\documentclass[prd,preprint,eqsecnum,nofootinbib,amsmath,amssymb,tightenlines,superscriptaddress]{revtex4}
\pdfoutput=1
\usepackage[pdftex]{graphicx}
\usepackage{epstopdf}
\usepackage{bm}
\usepackage{amsfonts}
\usepackage{amssymb}
\usepackage{slashbox}

\usepackage{dsfont}

\def\b{{\bm b}}

\def\B{{\bm B}}
\def\C{{\bm C}}

\def\bcalB{{\bm{\mathcal B}}}
\def\bcalP{{\bm{\mathcal P}}}

\def\uOmega{\underline{\Omega}}
\def\Ybar{\overline{Y}}

\def\Nc{N_{\rm c}}
\def\CA{C_{\rm A}}
\def\dA{d_{\rm A}}

\def\alphas{\alpha_{\rm s}}

\def\Re{\operatorname{Re}}

\def\sh{\operatorname{sh}}

\def\grad{{\bm\nabla}}

\def\ix{{\rm i}}
\def\fx{{\rm f}}
\def\xx{{\rm x}}
\def\xbx{{\bar{\rm x}}}
\def\yx{{\rm y}}
\def\ybx{{\bar{\rm y}}}
\def\zx{{\rm z}}
\def\Bx{\xbx}
\def\Ax{\ybx}
\def\bx{\yx}

\def\Hilbert{{\mathbb H}}

\def\tildeV{{\widetilde V}}

\def\yfrak{{\mathfrak y}}

\def\seq{{\rm seq}}

\def\four{{(4)}}
\def\fourb{{(\bar 4)}}
\def\ff{{(44)}}
\def\ffb{{(4\bar 4)}}

\begin {document}



\title
    {
      The LPM effect in sequential bremsstrahlung: \\
      4-gluon vertices
    }

\author{Peter Arnold}
\affiliation
    {%
    Department of Physics,
    University of Virginia,
    Charlottesville, Virginia 22904-4714, USA
    \medskip
    }%
\author{Han-Chih Chang}
\affiliation
    {%
    Department of Physics,
    University of Virginia,
    Charlottesville, Virginia 22904-4714, USA
    \medskip
    }%
\author{Shahin Iqbal}
\affiliation
    {%
    National Centre for Physics, \\
    Quaid-i-Azam University Campus,
    Islamabad, 45320 Pakistan
    \medskip
    }%

\date {\today}

\begin {abstract}%
{%
The splitting processes of bremsstrahlung and pair production in a medium
are coherent over large distances in the very high energy limit,
which leads to a suppression known as the Landau-Pomeranchuk-Migdal
(LPM) effect.  In this paper, we continue study of the case when the coherence
lengths of two consecutive splitting processes overlap (which is
important for understanding corrections to standard treatments
of the LPM effect in QCD), avoiding soft-gluon approximations.
In particular, this paper completes the calculation of
the rate for real double gluon bremsstrahlung
from an initial gluon
with various simplifying assumptions
(thick media; $\hat q$ approximation; and large $\Nc$)
by now including processes involving
4-gluon vertices.
}%
\end {abstract}

\maketitle
\thispagestyle {empty}

{\def\boldmath{}\tableofcontents}
\newpage


\section{Introduction and Result}
\label{sec:intro}

When passing through matter, high energy particles lose energy by
showering, via the splitting processes of hard bremsstrahlung and pair
production.  At very high energy, the quantum mechanical duration of
each splitting process, known as the formation time, exceeds the mean
free time for collisions with the medium, leading to a significant
reduction in the splitting rate known as the Landau-Pomeranchuk-Migdal
(LPM) effect \cite{LP,Migdal}.
A long-standing problem in field theory has
been to understand how to implement this effect in cases where
the formation times of two consecutive splittings overlap.

Let $x$ and $y$ be the longitudinal
momentum fractions of two consecutive bremsstrahlung gauge bosons.
In the limit $y \ll x \ll 1$, the problem of overlapping formation
times has been analyzed at leading logarithm order in
refs.\ \cite{Blaizot,Iancu,Wu}
in the context of
energy loss of high-momentum partons traversing
a QCD medium (such as a quark-gluon plasma).
We subsequently developed and implemented field theory
formalism needed for the more general case where $x$ and $y$ are
arbitrary \cite{2brem,seq,dimreg}.
In this paper, we finally complete the calculation of the effect
of overlapping formation times on the differential rate
$d\Gamma/dx\,dy$ for double bremsstrahlung
from an initial high-energy gluon
(with various simplifying assumptions detailed below).
The missing element, presented in this paper, is the inclusion
of processes involving the 4-gluon vertex.


\subsection {What we compute (and what we do not)}
\label {eq:what}

\begin {figure}[t]
\begin {center}
  \includegraphics[scale=0.5]{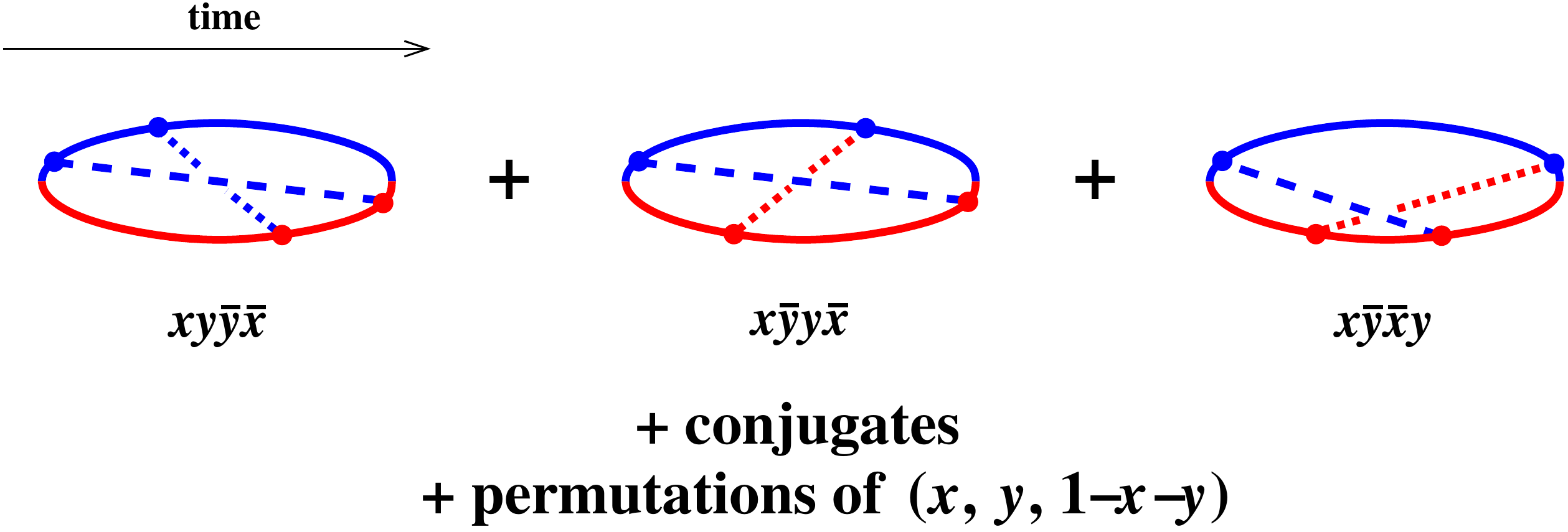}
  \caption{
     \label{fig:subset2}
     The subset of interference contributions to double splitting
     previously evaluated in ref.\ \cite{2brem},
     the ``crossed'' diagrams,
     depicted as
     amplitudes (blue) sewn together with conjugate amplitudes (red).
     The dashed lines are colored according to whether they
     were first emitted in the amplitude or conjugate amplitude.
     To simplify the drawing, all particles, including
     bremsstrahlung gluons, are indicated by straight or curved lines.
     The long-dashed and short-dashed lines
     are the daughters with momentum fractions $x$ and $y$
     respectively. 
     The naming of the diagrams indicates the time order
     in which emissions occur in the amplitude and conjugate amplitude.
     For instance, $x\bar y y \bar x$ means first
     (i) $x$ emission in the amplitude, then (ii) $y$ emission in the
     conjugate amplitude, then (iii) $y$ emission in the amplitude,
     and then (iv) $x$ emission in the conjugate amplitude.
  }
\end {center}
\end {figure}

\begin {figure}[t]
\begin {center}
  \includegraphics[scale=0.5]{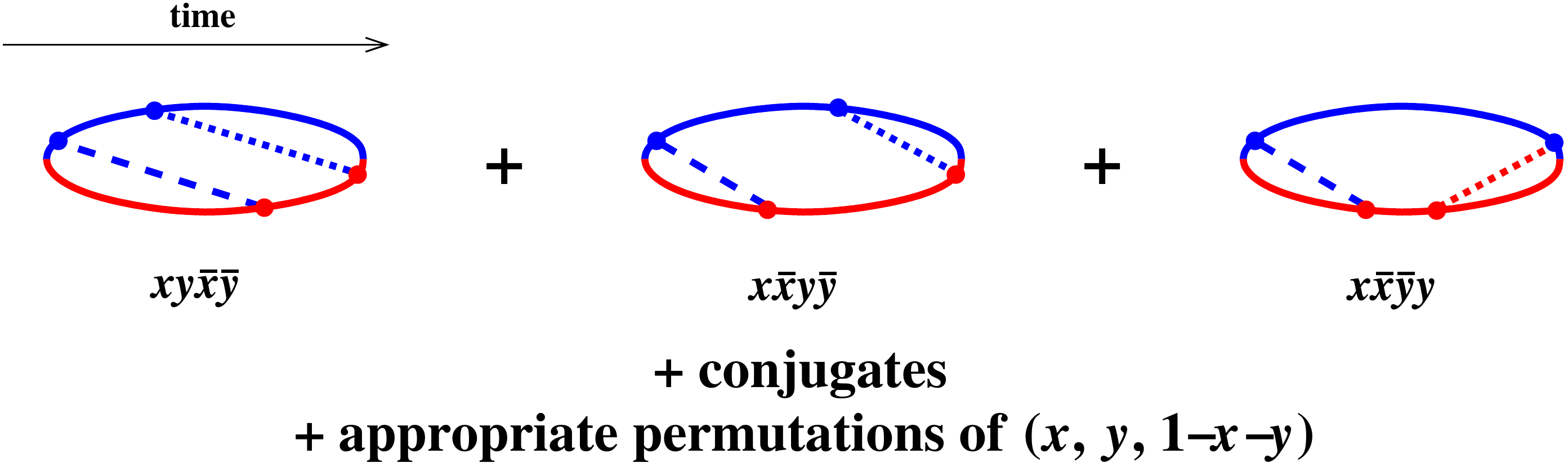}
  \caption{
     \label{fig:seqs2}
     The interference contributions evaluated in ref.\ \cite{seq}:
     the ``sequential'' diagrams.
  }
\end {center}
\end {figure}

The preceding work \cite{2brem,seq,dimreg} computed all of
the interference contributions involving only 3-gluon vertices,
which are presented by the diagrams of figs.\ \ref{fig:subset2}
and \ref{fig:seqs2}, which we respectively refer to as
``crossed'' and ``sequential'' diagrams.
The upper (blue) part of each diagram
depicts a contribution to the amplitude and the lower (red) part
depicts a contribution to the conjugate amplitude.
Only the high energy particles are shown;
their (many) interactions with the medium are implicit.
(See ref.\ \cite{2brem} for more details.)

\begin {figure}[t]
\begin {center}
  \includegraphics[scale=0.5]{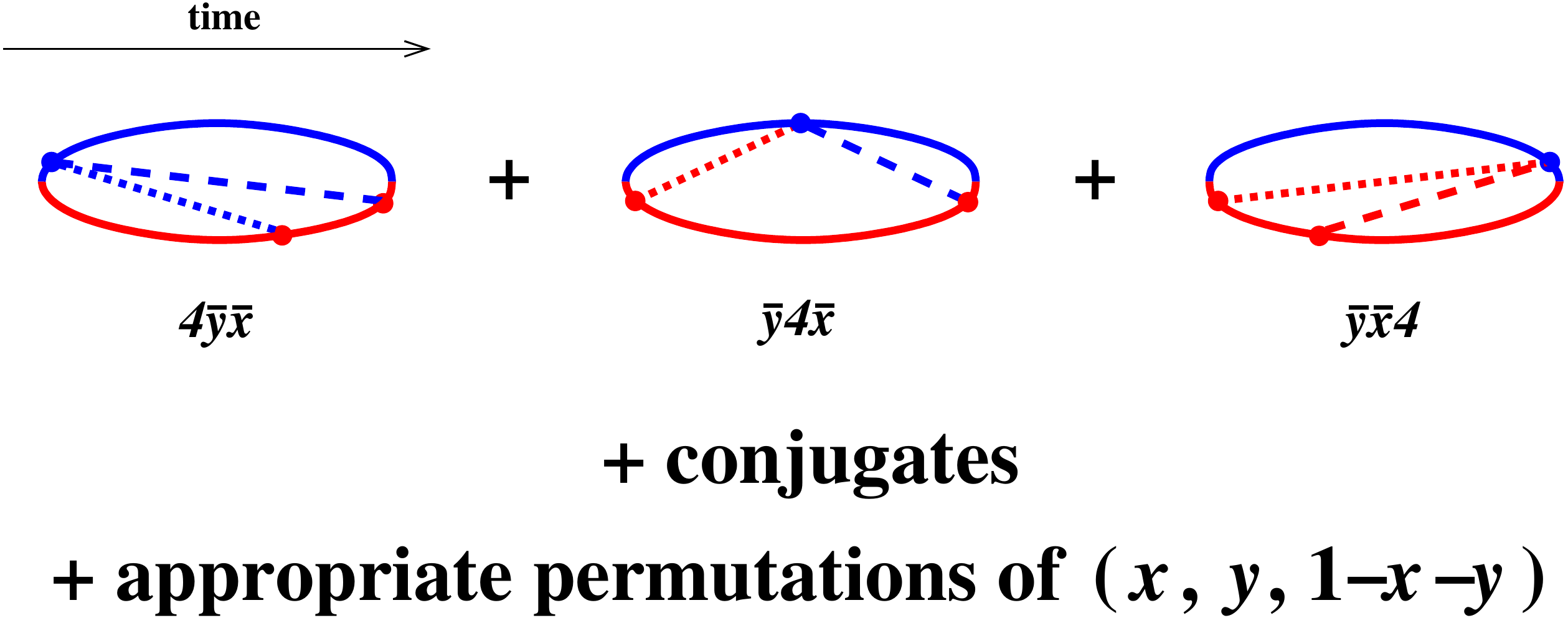}
  \caption{
     \label{fig:4subset2}
     The interference contributions involving a single 4-gluon
     vertex.  The naming conventions are the same as described in the
     caption of fig.\ \ref{fig:subset2} with the addition that
    ``4'' indicates
     a 4-gluon vertex where $x$ and $y$ are emitted simultaneously.
  }
\end {center}
\end {figure}

\begin {figure}[t]
\begin {center}
  \includegraphics[scale=0.5]{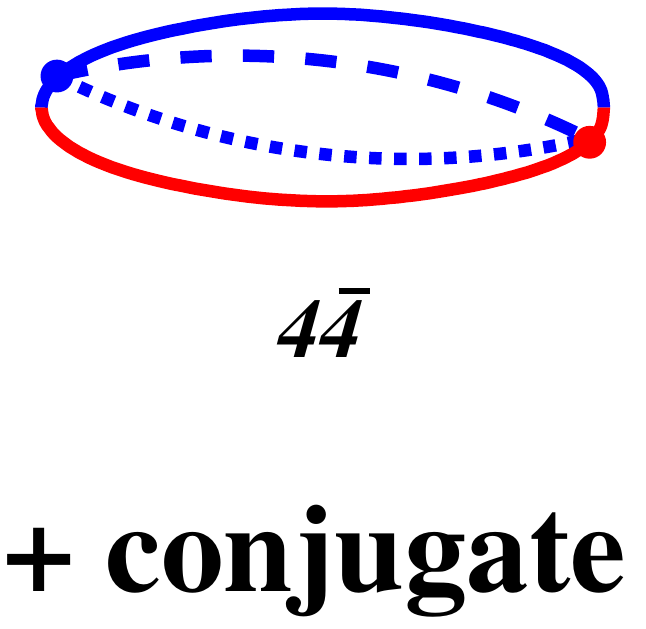}
  \caption{
     \label{fig:44subset2}
     The interference contribution involving two 4-gluon
     vertices.
  }
\end {center}
\end {figure}

In this paper, we will evaluate the remaining contributions, which
are the diagrams involving 4-point gluon vertices, shown in
figs.\ \ref{fig:4subset2} and \ref{fig:44subset2}.
(We will see later, by a symmetry argument,
that the $\bar y 4 \bar x$ contribution in fig.\ \ref{fig:4subset2}
vanishes.)
Once we find the correct normalization of the 4-gluon vertex in
our formalism, the evaluation of these diagrams will be a
relatively straightforward application of techniques developed in
previous papers \cite{2brem,seq}.

As discussed in the preceding work \cite{2brem,seq}, it is possible to set
up the formalism in a quite general way that would require both
highly non-trivial numerics and a non-trivial treatment of color dynamics
to implement, but
one can proceed much further analytically by making a few
additional approximations.  Though the methods we discuss in this
paper can be applied more generally, we will follow refs.\ \cite{2brem,seq}
when it comes to explicit calculations, by making the following approximations.
\begin {itemize}
\item
  We will assume that the medium is static, uniform and infinite (which
  in physical terms means approximately uniform over the formation time
  and corresponding formation length).
\item
  We take the large-$\Nc$ limit of QCD to simplify the color dynamics.
\goodbreak 
\item
  We make the multiple-scattering approximation to interactions with
  the medium, appropriate for very high energies and also known as the
  harmonic oscillator or $\hat q$ approximation.
\end {itemize}

In this paper, we focus on completing the calculation of
the rate for producing
two {\it real}\/ bremsstrahlung gluons ($g \to ggg$).
We defer to another time the related calculation of
the change in the {\it single}-bremsstrahlung rate
due to virtual corrections.  (In the special limiting
case $y \ll x \ll 1$, the sum of these real and virtual
processes has been worked out
in the context of leading parton
average energy loss in refs.\ \cite{Blaizot,Iancu,Wu} and is related
to anomalous scaling of the effective medium parameter $\hat q$
with energy.)

Finally, as discussed in ref.\ \cite{seq}, the double bremsstrahlung
rate $d\Gamma/dx\,dy$ by itself includes processes where
two single-bremsstrahlung processes are separated by times large
compared to their corresponding formation times.
In the idealization of an infinite, uniform medium, this
causes $d\Gamma/dx\,dy$ to be formally infinite.  But what
we actually want to know is the {\it correction} to
double bremsstrahlung
due to overlapping formation times,
\begin {equation}
   \Delta\, \frac{d\Gamma}{dx\,dy} \equiv
   \frac{d\Gamma}{dx\,dy}
   -
   \left[ \frac{d\Gamma}{dx\,dy} \right]_{\rm IMC} ,
\label{eq:DeltaDef}
\end {equation}
where $[d\Gamma/dx\,dy]_{\rm IMC}$ represents the
idealized in-medium
``Monte Carlo'' result one would obtain based only on the rates
for single-bremsstrahlung processes.  See the introduction of
ref.\ \cite{seq} for a detailed explanation.
The correction $\Delta\,d\Gamma/dx\,dy$ is finite and
only depends on time separations that are $\lesssim$ formation times.
The subtraction (\ref{eq:DeltaDef}) is an issue
relevant only to the the sequential diagram
contributions of fig.\ \ref{fig:seqs2}; we will not need to
worry about it when evaluating the 4-gluon vertex diagrams of
figs.\ \ref{fig:4subset2} and \ref{fig:44subset2}.
The subtraction will be relevant only for presenting complete, final
results for the double bremsstrahlung rate, which combine all
the contributions of figs.\ \ref{fig:subset2}--\ref{fig:44subset2}.


\subsection {Preview of Results}

\begin {figure}[t]
\begin {center}
  \includegraphics[scale=0.7]{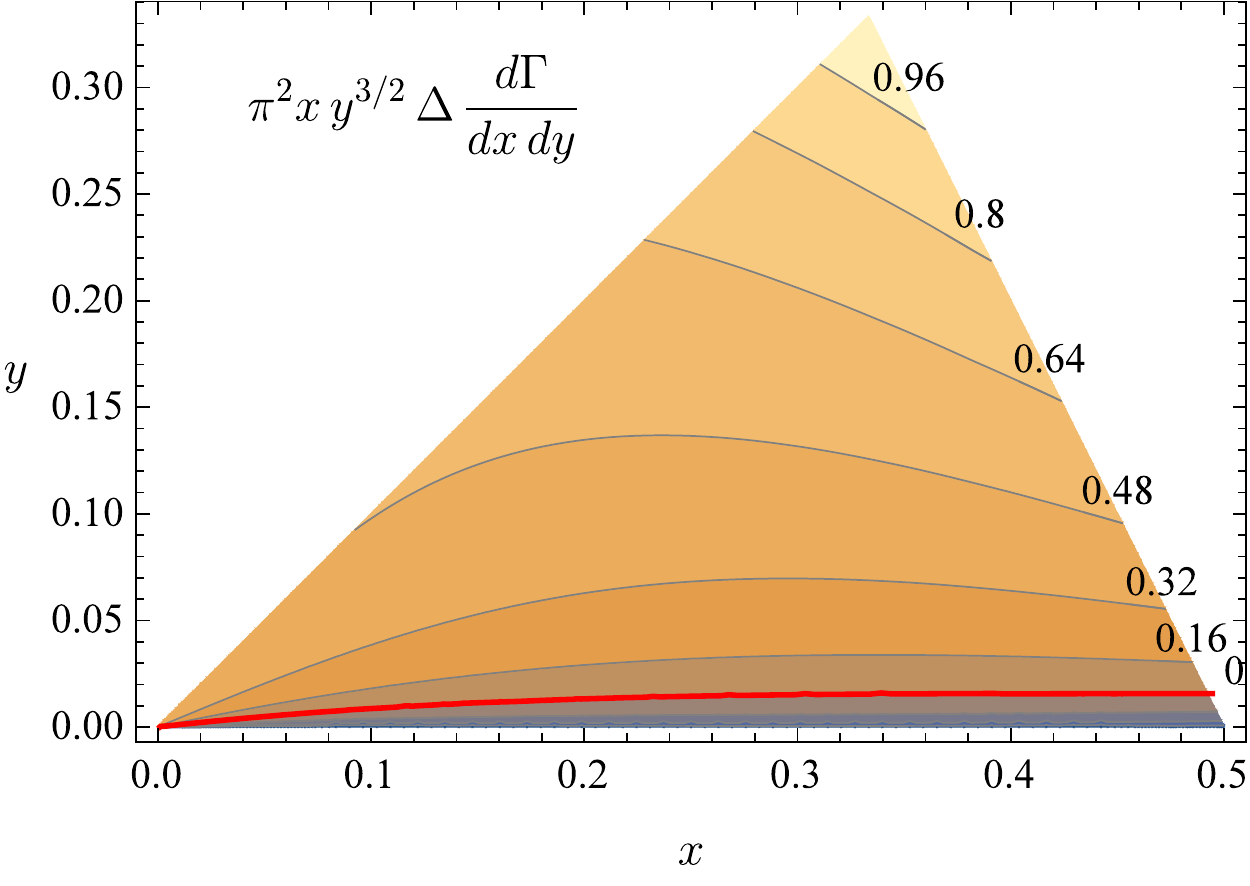}
  \caption{
     \label{fig:4pointResult}
     Result for $\pi^2 x y^{3/2} \, \Delta\, d\Gamma/dx\,dy$ in units
     of $\CA^2\alphas^2 \sqrt{\hat q_{\rm A}/E}$ [which is equivalent
     to saying the result for $\Delta\, d\Gamma/dx\,dy$ in units of
     (\ref{eq:units})].
     Since all three final state particles are gluons and so are identical
     particles, we only show results
     for the region $y < x < z \equiv 1{-}x{-}y$.
     (All other orderings are related by permutation.)
     The red line shows where the
     result vanishes, dividing the sub-region of positive corrections from
     the sub-region of negative corrections.
     At the apex ($x{=}y{=}\frac13$) of the triangular region,
     $\pi^2 x y^{3/2} \,\Delta\,d\Gamma/dx\,dy =
      1.12 \, \CA^2 \alphas^2 \sqrt{\hat q_{\rm A}/E}$.
  }
\end {center}
\end {figure}

Numerical results for the total $\Delta\,d\Gamma/dx\,dy$ are shown in fig.\
\ref{fig:4pointResult}, which includes all contributions from
figs.\ \ref{fig:subset2}--\ref{fig:44subset2}.
In ref.\ \cite{seq}, it was shown that the contribution from
crossed and sequential diagrams (figs.\ \ref{fig:subset2} and \ref{fig:seqs2})
scale as $1/x y^{3/2}$ for $y \ll x \ll 1$, and for this reason
it has been convenient to show the result in fig.\ \ref{fig:4pointResult}
in units of
\begin {equation}
  \frac{\CA^2\alphas^2}{\pi^2 x y^{3/2}} \sqrt{\frac{\hat q_{\rm A}}{E}} .
\label {eq:units}
\end {equation}

\begin {figure}[t]
\begin {center}
  \includegraphics[scale=0.7]{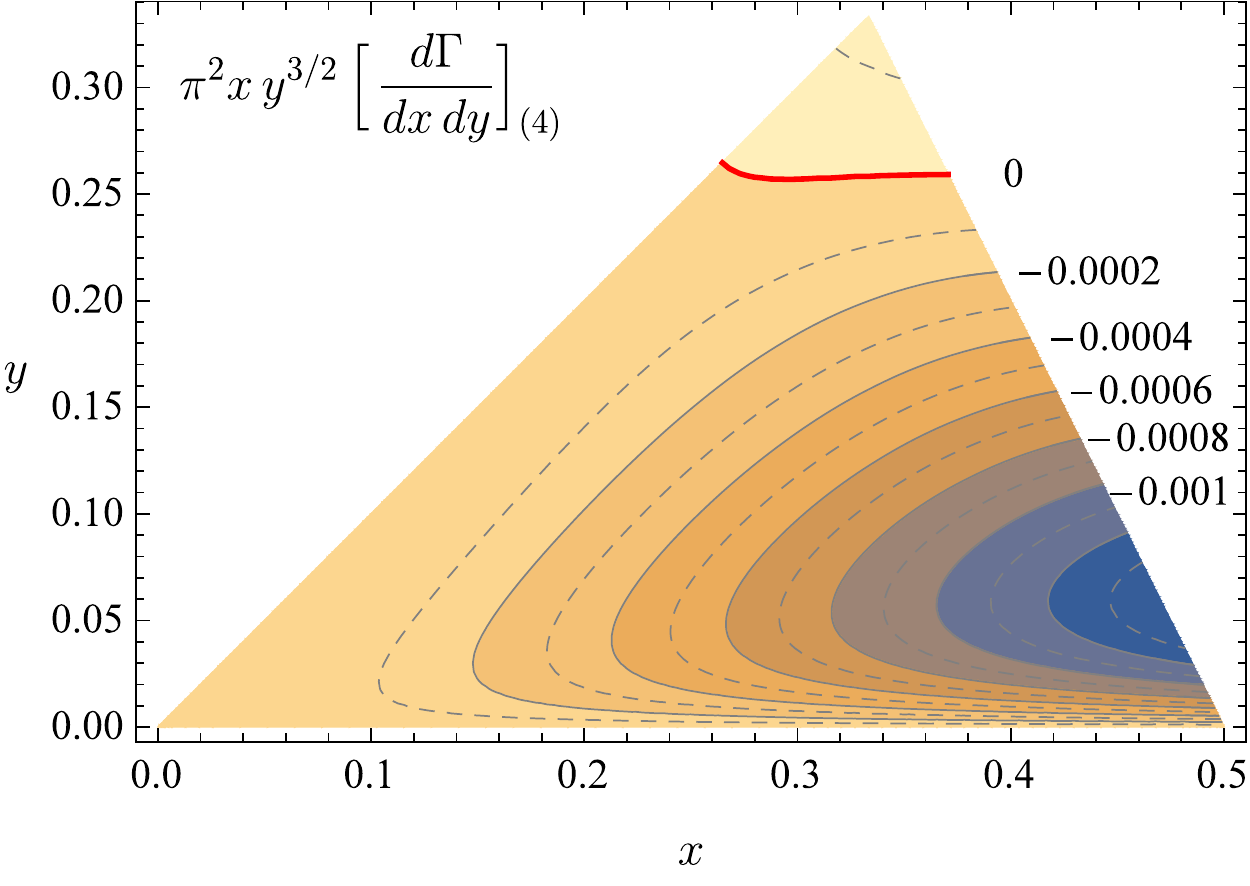}
  \caption{
     \label{fig:4pointResult4}
     As fig.\ \ref{fig:4pointResult} but only showing the contribution
     from the diagrams of fig.\ \ref{fig:4subset2}, which are the diagrams
     with a single 4-gluon vertex.
     At the apex,
     $\pi^2 x y^{3/2} [d\Gamma/dx\,dy]_\four =
      0.00012 \, \CA^2 \alphas^2 \sqrt{\hat q_{\rm A}/E}$.
  }
\end {center}
\end {figure}

\begin {figure}[t]
\begin {center}
  \includegraphics[scale=0.7]{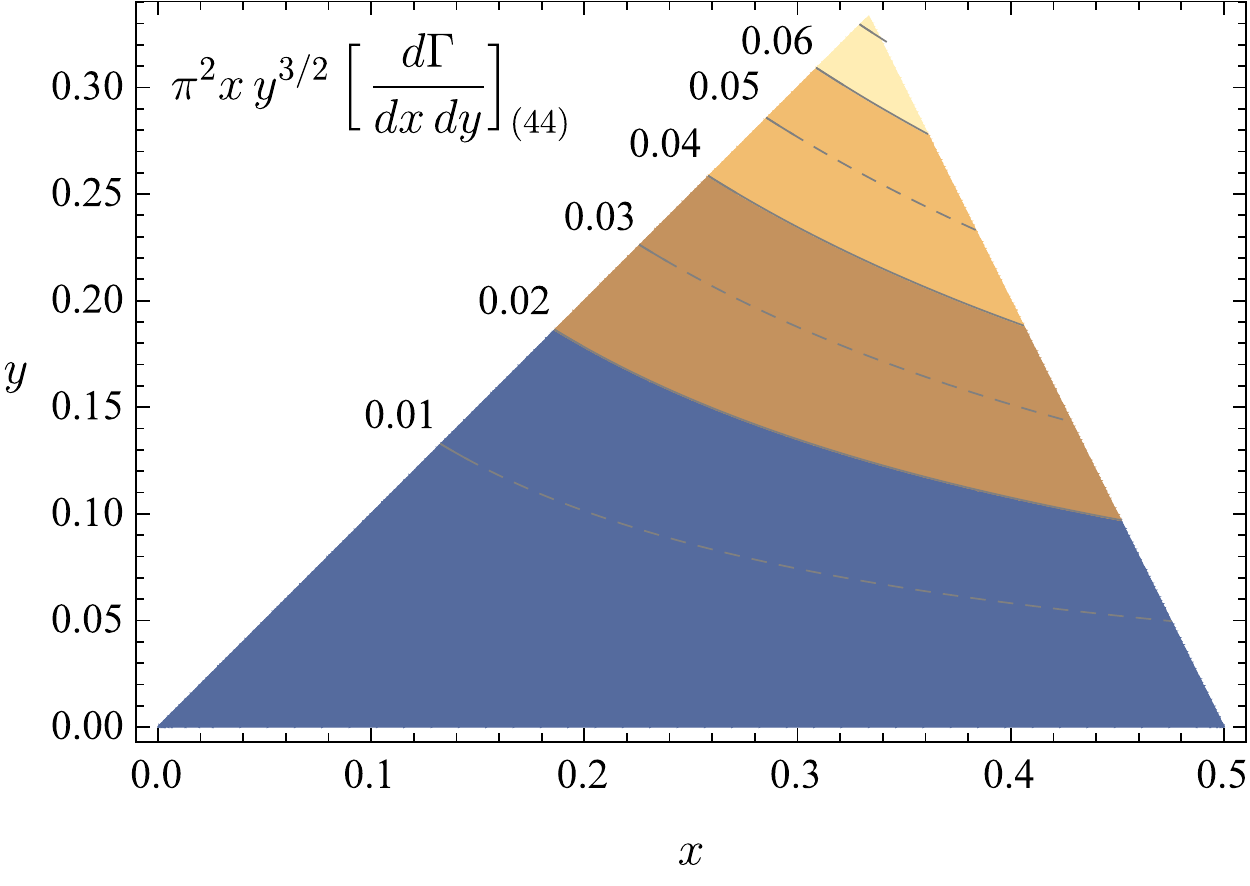}
  \caption{
     \label{fig:4pointResult44}
     As fig.\ \ref{fig:4pointResult} but only showing the contribution
     from the diagrams of fig.\ \ref{fig:44subset2}, which are the
     diagrams with two 4-gluon vertices.
     At the apex,
     $\pi^2 x y^{3/2} [d\Gamma/dx\,dy]_\ff =
      0.072 \, \CA^2 \alphas^2 \sqrt{\hat q_{\rm A}/E}$.
  }
\end {center}
\end {figure}

In comparison to the similar plot in ref.\ \cite{seq},
not much has changed: the inclusion of the
4-gluon vertex contributions of
figs.\ \ref{fig:4subset2} and \ref{fig:44subset2} in this paper have
had only a small effect on the total.  We show the contributions
of figs.\ \ref{fig:4subset2} and \ref{fig:44subset2} individually
in figs.\ \ref{fig:4pointResult4} and \ref{fig:4pointResult44}.
The first of these is numerically negligible compared to the total
of fig.\ \ref{fig:4pointResult}.  (We do not know any qualitative
explanation for why it should be so small.%
\footnote{
  Some readers may wonder if (i) this contribution vanishes for some
  unidentified reason
  and (ii) the small numbers are just artifacts of imprecise numerical
  calculations.  However, we have checked that fig.\ \ref{fig:4pointResult}
  does not change when we steadily increase the precision of our
  calculations (including the working precision of intermediate calculations).
}%
)
The second (fig.\ \ref{fig:4pointResult44})
is only a very modest contribution to the total.

None of the new, 4-gluon vertex contributions to $\Delta\,d\Gamma/dx\,dy$
grow as quickly as (\ref{eq:units}) for $y \ll x \ll 1$.
We find that they instead scale as $1/y^{1/2}$ in this limit.



\subsection {Outline and Referencing}

In the next section, we show how to calculate the $4\bar y\bar x$
interference diagram of fig.\ \ref{fig:4subset2}, which will be our
canonical example in this paper.
Section \ref{sec:other} then explains how to obtain all of
the other diagrams involving 4-gluon vertices.
A summary of final formulas is given in section \ref{sec:summary},
and we offer our brief conclusion in section \ref{sec:conclusion}.
Along the way, some details and cross-checks are relegated to
appendices.
In particular, for the sake of completeness, we
have collected in Appendix \ref{app:crossseq} the formulas
for crossed and sequential diagrams from refs.\ \cite{2brem,dimreg,seq},
so that this paper contains, in one place, all the formulas necessary
for implementing the complete calculation of $\Delta\,d\Gamma/dx\,dy$.
Also, the integral formula we will derive for $\Delta\,d\Gamma/dx\,dy$ is
a complicated expression that is painstaking to implement.  In
Appendix \ref{app:approx}, we provide, as an alternative,
a relatively simple analytic formula that has been fitted to approximate
fig.\ \ref{fig:4pointResult} very well.

In this paper, we will occasionally (in footnotes and appendices)
use the author acronym AI as shorthand for
Arnold and Iqbal \cite{2brem} so that, for example,
we may write ``AI (5.2)'' to refer to
eq.\ (5.2) of ref.\ \cite{2brem}.


\section {The \boldmath$4\bar y\bar x$ diagram}
\label {sec:4yx}

\subsection {Starting point}

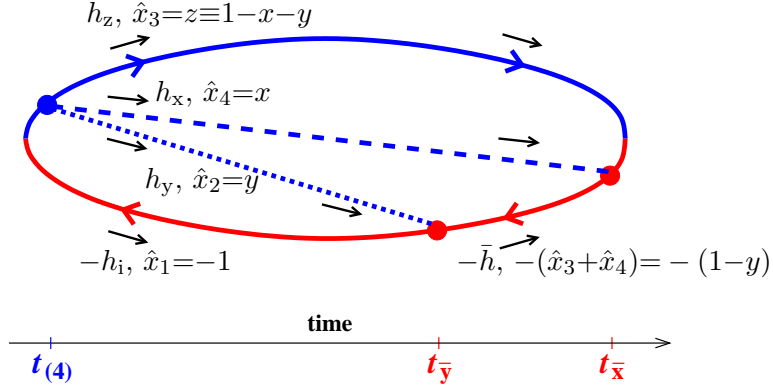
\begin {figure}[t]
\begin {center}
  \begin{picture}(250,150)(0,0)
  \put(0,0){\includegraphics[scale=0.5]{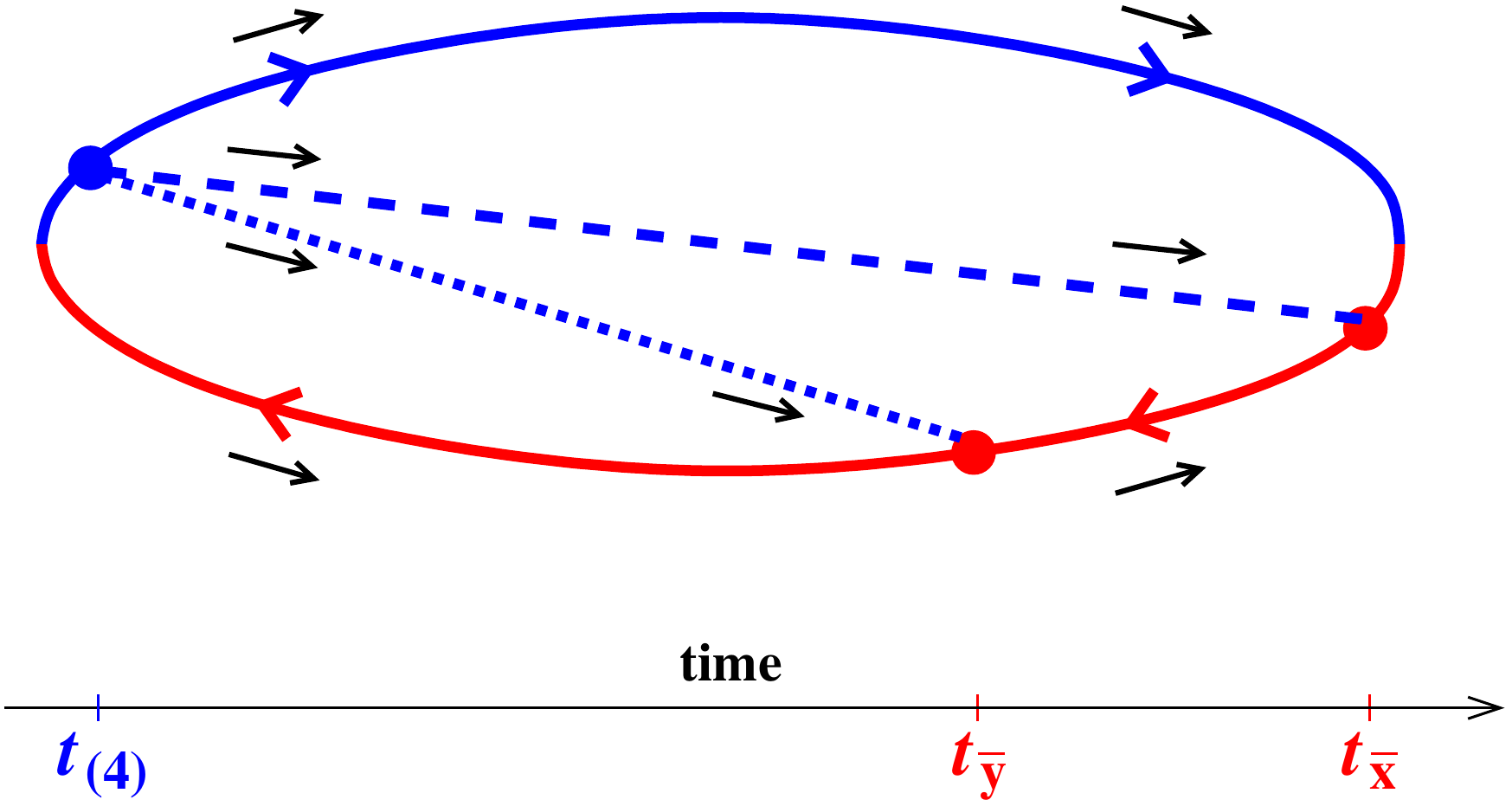}}
  \put(27,44){${-}h_{\rm i},\,\hat x_1{=}{-}1$}
  \put(52,75){$h_{\rm y},\,\hat x_2{=}y$}
  \put(56,108){$h_{\rm x},\,\hat x_4{=}x$}
  \put(30,138){$h_{\rm z},\,\hat x_3{=}z{\equiv}1{-}x{-}y$}
  \put(170,44){${-}\bar h,\,{-}(\hat x_3{+}\hat x_4){=}-(1{-}y)$}
  \end{picture}
  \caption{
     \label{fig:4yx}
     Labeling conventions for
     helicities $h_i$ and longitudinal momenta $x_i$
     for the $4\bar y\bar x$ interference diagram.
  }
\end {center}
\end {figure}

We start with the $4\bar y\bar x$ diagram shown in
fig.\ \ref{fig:4yx}.
In the notation of ref.\ \cite{2brem}, this is
\begin {align}
   \left[\frac{dI}{dx\,dy}\right]_{4\bar y\bar x}
   =
   \left( \frac{E}{2\pi} \right)^2
   \int_{t_\four < t_\ybx < t_\xbx}
   &
   \sum_{\rm pol.}
   \langle|i\,\overline{\delta H}|\B^\Bx\rangle \,
   \langle\B^\Bx,t_\Bx|\B^\Ax,t_\Ax\rangle
   \langle\B^\Ax|i\,\overline{\delta H}|\C_{34}^\Ax,\C_{12}^\Ax\rangle \,
\nonumber\\ &\times
   \langle\C_{34}^\ybx,\C_{12}^\ybx,t_\ybx|\C_{34}^\four,\C_{12}^\four,t_\four\rangle
   \langle\C_{34}^\four,\C_{12}^\four|{-}i\,\delta H|\rangle .
\label {eq:4yxstart}
\end {align}
$\langle\C_{34}^\ybx,\C_{12}^\ybx,t_\ybx|\C_{34}^\four,\C_{12}^\four,t_\four\rangle$
and
$\langle\B^\xbx,t_\xbx|\B^\ybx,t_\ybx\rangle$
represent, respectively, the
(i) 4-particle
evolution in the initial time interval $t_\four < t < t_\ybx$
in the figure, and
(ii) 3-particle evolution of the system in the final interval
$t_\ybx < t < t_\xbx$.  Because of the symmetries of the problem,
these have been reduced to effective (i) 2-particle
and (ii) 1-particle problems in non-Hermitian two-dimensional quantum mechanics,
described by effective transverse coordinates (i) $(\C_{34},\C_{12})$
and (ii) $\B$.
$\delta H$ represents the piece of the fundamental QCD Hamiltonian
associated with the splitting vertices for the high-energy particles
(as opposed to the interactions of those high-energy
particle with the medium, or the interaction of the medium with
itself).  So $\langle\C_{34}^\four,\C_{12}^\four|\delta H|\rangle$
represents the matrix element for the 4-gluon splitting vertex
in fig.\ \ref{fig:4yx}, appropriately normalized according to
the normalization conventions for the states
$|\C_{34}^\four,\C_{12}^\four\rangle$ and $|\rangle$ given
in ref.\ \cite{2brem}.

Above, $\C_{ij} \equiv (\b_i-\b_j)/(x_i+x_j)$ where the $\b_i$ are
the various transverse positions of the individual particles and
$x_i$ are their longitudinal momentum fractions (defined as negative
for particles in the conjugate amplitude). 
$\B \equiv \B_{12} = \B_{23} = \B_{31}$ is defined similarly for
the case of three particles.

\begin {figure}[t]
\begin {center}
  \begin{picture}(400,110)(0,0)
  \put(25,20){\includegraphics[scale=0.7]{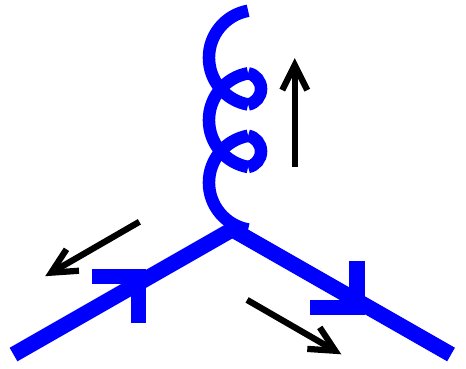}}
  \put(5,10){$\b_i,a_i$}
  \put(115,10){$\b_j,a_j$}
  \put(60,100){$\b_k,a_k$}
  \put(20,48){$x_i,h_i$}
  \put(70,15){$x_j,h_j$}
  \put(88,70){$x_k,h_k$}
  \put(130,55){$\displaystyle{
       = \quad - \frac{g (T_R^{a_k})_{a_j,a_i}}{2 E^{3/2}}
       \> \bcalP_{h_i h_j h_k}\!(x_i,x_j,x_k) \cdot
       \grad \delta^{(2)}({\bcalB}_{ji})
    }$}
  \end{picture}
  \\[10pt]
  \begin{picture}(400,130)(0,0)
  \put(25,2){\includegraphics[scale=0.7]{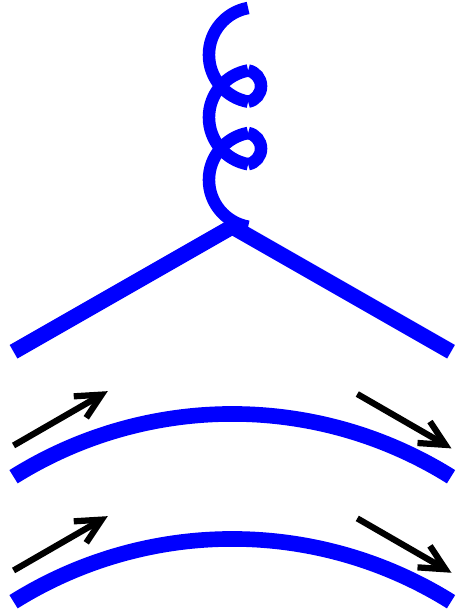}}
  \put(11,0){$\b_n$}
  \put(120,0){$\b_s$}
  \put(11,26){$\b_m$}
  \put(120,26){$\b_r$}
  \put(50,20){$x_n$}
  \put(84,20){$x_s$}
  \put(50,46){$x_m$}
  \put(84,46){$x_r$}
  \put(130,55){$\displaystyle{=}$}
  \put(150,28){\includegraphics[scale=0.7]{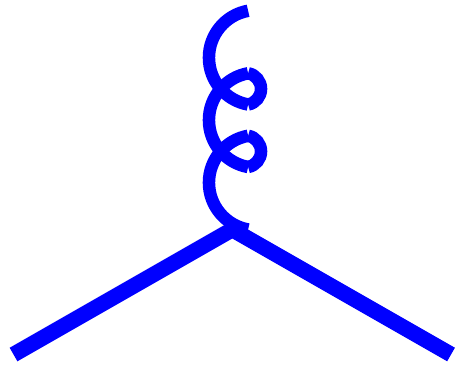}}
  \put(250,55){$\displaystyle{
       \times ~~
       |x_m+x_n|^{-1} \, \delta^{(2)}({\bcalB}_{mn} {-} {\bcalB}_{rs})
    }$}
  \end{picture}
  \caption{
     \label{fig:dH}
     The diagrammatic rules for splittings linking (via
     either $-i\,\delta H$ or $+i\,\overline{\delta H}$)
     the state
     $|\rangle$ to $|\B\rangle$ (top rule) or $|\B\rangle$ to
     $|\C_{34},\C_{12}\rangle$ or permutation thereof (bottom rule). 
     ${\bcalB}_{uv} \equiv (\b_u-\b_v)/(x_u+x_v)$ and may refer, in
     different contexts, to $\pm$ the 3-particle $\B$, or one
     of the 4-particle $\C_{uv}$, or to some mixture.  However,
     note that
     ${\bcalB}_{ji} = {\bcalB}_{kj} = {\bcalB}_{ik}$
     in the top rule, which can be used
     to always write expressions in terms of 3-particle $\B$ and/or
     4-particle $\C_{ij}$'s.
     The blue arrows on the particle line indicate color flow of
     color representation $R$.  (In the case of $R{=}{\rm A}$,
     appropriate to $g \to gg$ splitting, the direction of the color
     flow does not matter.)  $\b_l$, $a_l$, $x_l$, and $h_l$ indicate the
     transverse position, color index, longitudinal momentum, and
     helicity of each particle.  The black arrows give the
     convention for the flow of $x_l$ and $h_l$ in the
     statement of the rule, and these values should be negated if
     they are instead defined by flow in the opposite direction.
     In the bottom rule, color and helicity
     indices and their contractions are not explicitly
     shown for the spectators because they are trivially contracted.
     Conservation of longitudinal momentum means $x_i+x_j+x_k=0$ (top)
     and additionally $x_m=x_r$ and $x_n = x_s$ (bottom).
  }
\end {center}
\end {figure}
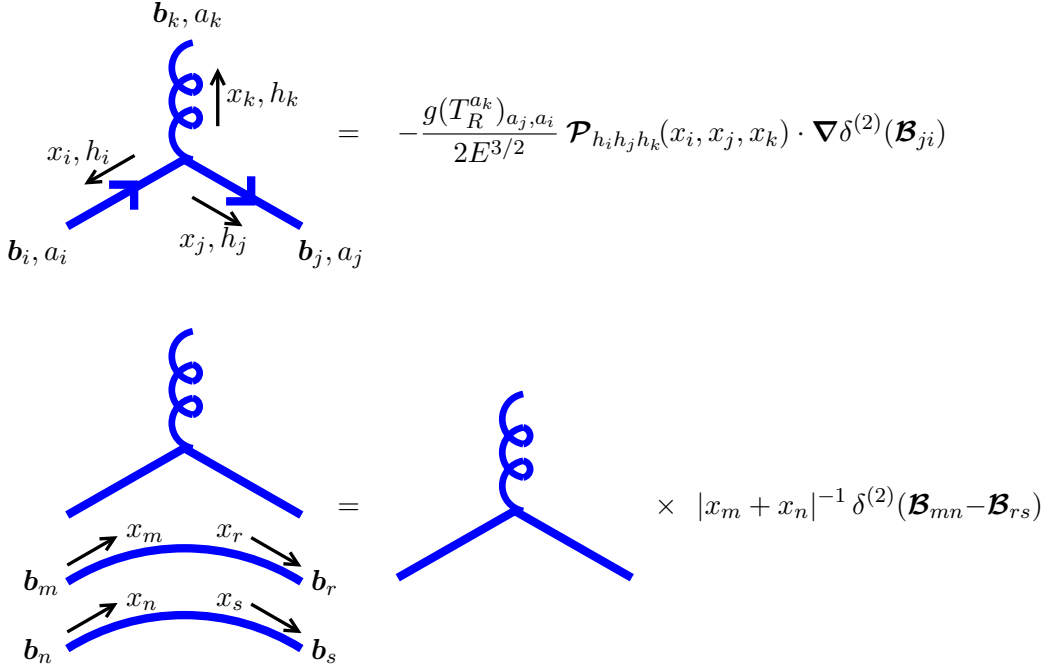

The appropriately normalized results for the 3-gluon vertices were
found in ref.\ \cite{2brem}:%
\footnote{
  AI (4.13--15)
}
\begin {equation}
   \langle \B | \delta H |\rangle
   = - \frac{i g T^{\rm color}_{i \to jk}}{2 E^{3/2}} \, \bcalP_{i \to jk}
     \cdot \grad \delta^{(2)}(\B)
\label {eq:dH32}
\end {equation}
and
\begin {align}
   \langle\C_{41},\C_{23}|\delta H|\B\rangle
   = - \frac{i g T^{\rm color}_{i \to jk}}{2 E^{3/2}} \, \bcalP_{i \to jk}
     \cdot \grad \delta^{(2)}(\C_{23}) \,
     |\hat x_4+\hat x_1|^{-1} \, \delta^{(2)}(\C_{41}-\B) ,
\label {eq:dH43}
\end {align}
where $T^{\rm color}$ are color generators
and the $\bcalP_{i\to jk}$ are proportional to square roots of helicity-dependent,
vacuum Dokshitzer-Gribov-Lipatov-Altarelli-Parisi (DGLAP) splitting
functions.
These were translated into the more general diagrammatic rules of
fig.\ \ref{fig:dH}, which apply to
$\langle \B |\, {-}i\,\delta H |\rangle$,
$\langle\C_{ij},\C_{kl}|\,{-}i\,\delta H|\B\rangle$,
$\langle|\, {-}i\,\delta H |\B\rangle$ and
$\langle\B|\,{-}i\,\delta H|\C_{ij},\C_{kl}\rangle$,
as well as similar matrix elements
$\langle\cdots|\,{+}i\,\overline{\delta H}|\cdots\rangle$ relevant
to evolution in the conjugate amplitude.  [The bar over
$\overline{\delta H}$ here and in formulas like (\ref{eq:4yxstart})
is just a notation for emphasizing that $\delta H$ is operating on
particles in the conjugate amplitude in those cases.]

In appendix \ref{app:4matrix}, we apply the same methodology to
evaluating the 4-gluon vertex we need above and find
\begin {align}
  \langle \C_{34},\C_{12} | \delta H |\rangle
  =
       g^2 & \Bigl[
           f^{a_1 a_2 e} f^{a_3 a_4 e}
             (\delta_{h_1,-h_3} \delta_{h_4,-h_2} - \delta_{h_1,-h_4} \delta_{h_2,-h_3})
\nonumber\\ & \quad
         + f^{a_1 a_3 e} f^{a_2 a_4 e}
             (\delta_{h_1,-h_2} \delta_{h_3,-h_4} - \delta_{h_1,-h_4} \delta_{h_2,-h_3})
\nonumber\\ & \quad
         + f^{a_1 a_4 e} f^{a_2 a_3 e}
             (\delta_{h_1,-h_2} \delta_{h_3,-h_4} - \delta_{h_1,-h_3} \delta_{h_4,-h_2})
       \Bigr]
\nonumber\\ &
       \times
       (2E)^{-2} |x_1 x_2 x_3 x_4|^{-1/2} |x_3+x_4|^{-1} \,
       \delta^{(2)}(\C_{12}) \, \delta^{(2)}(\C_{34}) ,
\label {eq:dHCC}
\end {align}
where $a_i$ and $h_i$ are the color index and helicity $\pm$ associated
with particle $i$.  The first few lines of (\ref{eq:dHCC}) can be
recognized as having the structure of the usual relativistic
Feynman rule for a 4-gluon vertex; the last line has the normalization
factors appropriate for the way we normalize
the transverse position
variables $\C_{ij}$ and the state $|\C_{34},\C_{12}\rangle$ \cite{2brem}.
The two delta functions in the last line,
$\delta^{(2)}(\C_{12}) \, \delta^{(2)}(\C_{34})
 \propto \delta^{(2)}(\b_1-\b_2) \, \delta^{(2)}(\b_3-\b_4)$,
enforce that the four particles
all be in the same place
($\b_1{=}\b_2{=}\b_3{=}\b_4$)
at the time of the 4-point interaction.
(Generically, two $\delta$-functions may seem insufficient to enforce this,
but in our problem the positions $\b_i$ are already implicitly constrained by
the additional condition
$x_1 \b_1 + x_2 \b_2 + x_3 \b_3 + x_4 \b_4 = 0$
with $x_1+x_2+x_3+x_4=0$.
See section III of ref.\ \cite{2brem}.)

\begin {figure}[t]
\begin {center}
  \begin{picture}(425,110)(0,10)
  \put(25,20){\includegraphics[scale=0.7]{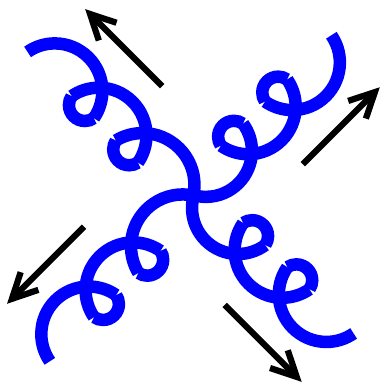}}
  \put(10,15){$\b_i,a_i$}
  \put(95,15){$\b_j,a_j$}
  \put(10,100){$\b_l,a_l$}
  \put(95,100){$\b_k,a_k$}
  \put(23,48){$h_i$}
  \put(70,18){$h_j$}
  \put(100,68){$h_k$}
  \put(50,95){$h_l$}
  \put(130,55){$\displaystyle{
     = \quad
     \begin {aligned}
       \mp i g^2 & \Bigl[
           f^{a_i a_j e} f^{a_k a_l e}
             (\delta_{h_i,-h_k} \delta_{h_l,-h_j} - \delta_{h_i,-h_l} \delta_{h_j,-h_k})
     \\ & \quad
         + f^{a_i a_k e} f^{a_j a_l e}
             (\delta_{h_i,-h_j} \delta_{h_k,-h_l} - \delta_{h_i,-h_l} \delta_{h_j,-h_k})
     \\ & \quad
         + f^{a_i a_l e} f^{a_j a_k e}
             (\delta_{h_i,-h_j} \delta_{h_k,-h_l} - \delta_{h_i,-h_k} \delta_{h_l,-h_j})
       \Bigr]
     \\ &
       \times
       (2E)^{-2} |x_i x_j x_k x_l|^{-1/2} |x_k+x_l|^{-1} \,
       \delta^{(2)}(\bcalB_{ij}) \, \delta^{(2)}(\bcalB_{kl})
     \end {aligned}
  }$}
  \end{picture}
  \caption{
     \label{fig:dH4point}
     The diagrammatic rule for a 4-gluon vertex
     without additional
     spectators (e.g.\ as in the $4\bar y\bar x$, $\bar y\bar x4$ and
     $4\bar 4$ diagrams of figs.\ \ref{fig:4subset2} and \ref{fig:44subset2}
     but not the $\bar y4\bar x$ diagram).
     The rule is symmetric
     under permutations of the four gluon lines, though this
     is not obvious from the way it is written.  The upper and lower
     signs of $\mp$ apply when the 4-gluon interaction is in the
     amplitude and conjugate amplitude, respectively.
  }
\end {center}
\end {figure}

A diagrammatic version of (\ref{eq:dHCC}) is given in
fig.\ \ref{fig:dH4point}.  Like the top graph of
fig.\ \ref{fig:dH}, this particular rule only applies when there
are no other particle lines present at that time.  So, it can
be used for $4\bar y\bar x$ and $\bar y\bar x 4$ in
fig.\ \ref{fig:4subset2} but not for $\bar y4\bar x$.
The 4-point vertex requires different normalization factors
in the latter case, which we give in Appendix \ref{app:BHB},
but that detail
is unimportant because $\bar y4\bar x$ turns out to vanish.

The sign $\mp$ in fig.\ \ref{fig:dH4point} simply reflects the fact
that in the amplitude the vertex corresponds to matrix elements of
$-i\,\delta H$ whereas in the conjugate amplitude it corresponds to
matrix elements of $+i\,\delta H$ (which we denote as
$+i\,\overline{\delta H}$).%
\footnote{
  Readers may wonder why there is not a similar explicit $\mp$
  sign in the 3-gluon vertex rule of fig.\ \ref{fig:dH}.
  The reason is because that sign is already there, hidden in the formulation
  of the rule.  As mentioned in the caption of fig.\ \ref{fig:dH},
  ${\bcalB}_{uv} \equiv (\b_u-\b_v)/(x_u+x_v)$.  However, our convention
  for momentum fractions $x_i$ is that they are negative for particles
  in the conjugate amplitude.  So, if going from consideration of the
  rule applied to splitting of particles in the amplitude to the same
  rule applied to splitting of particles in the conjugate amplitude,
  the value of the ${\bcalB_{ji}}$ will automatically negate.
  Since $\grad\delta^{(2)}(\bcalB_{ji})$ is an odd function of
  $\bcalB_{ji}$, this automatically takes care of the sign difference
  between $-i\,\delta H$ and $+i\,\overline{\delta H}$.
}


\subsection {Color routings}
\label {sec:color}

\begin {figure}[t]
\begin {center}
  \includegraphics[scale=0.8]{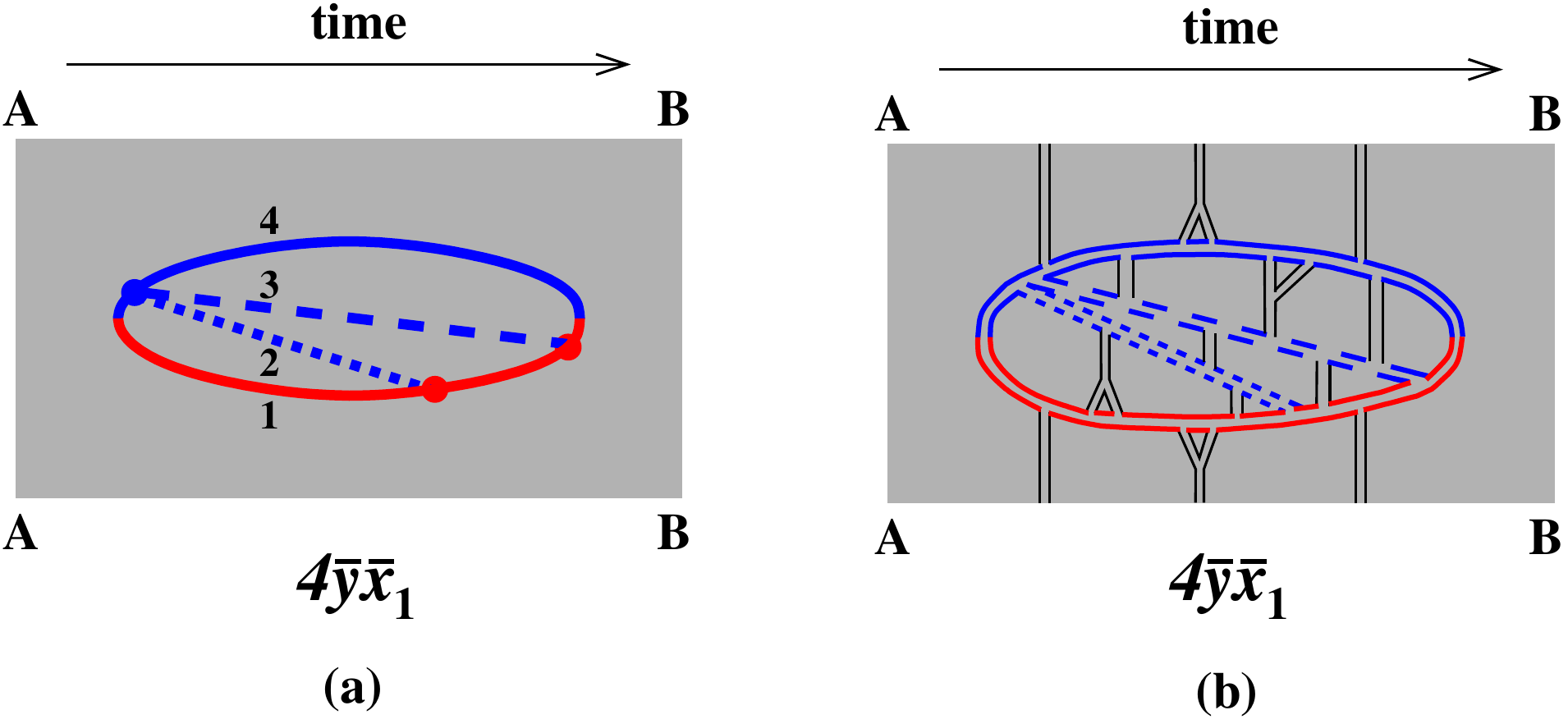}
  \caption{
     \label{fig:4yx1}
     One of the two distinct large-$\Nc$
     color routings of the $4\bar y\bar x$ interference
     diagram drawn on a cylinder
     (similar to fig.\ 23 of ref.\ \cite{seq},
     and following the general convention of refs.\ \cite{2brem,seq} for
     discussing time-ordered large-$\Nc$ planar diagrams).
     The top edge AB of the shaded region is to be identified with the
     bottom edge AB.
     (b) explicitly shows the corresponding color flow for an
     example of
     medium background field correlations (black) that gives a planar
     diagram (and so leading-order in $1/\Nc$).
     In our notation, this interference contribution could be referred
     to as either $4\bar y\bar x_1$ or $4\bar y\bar z_2$.
  }
\end {center}
\end {figure}

\begin {figure}[t]
\begin {center}
 \includegraphics[scale=0.8]{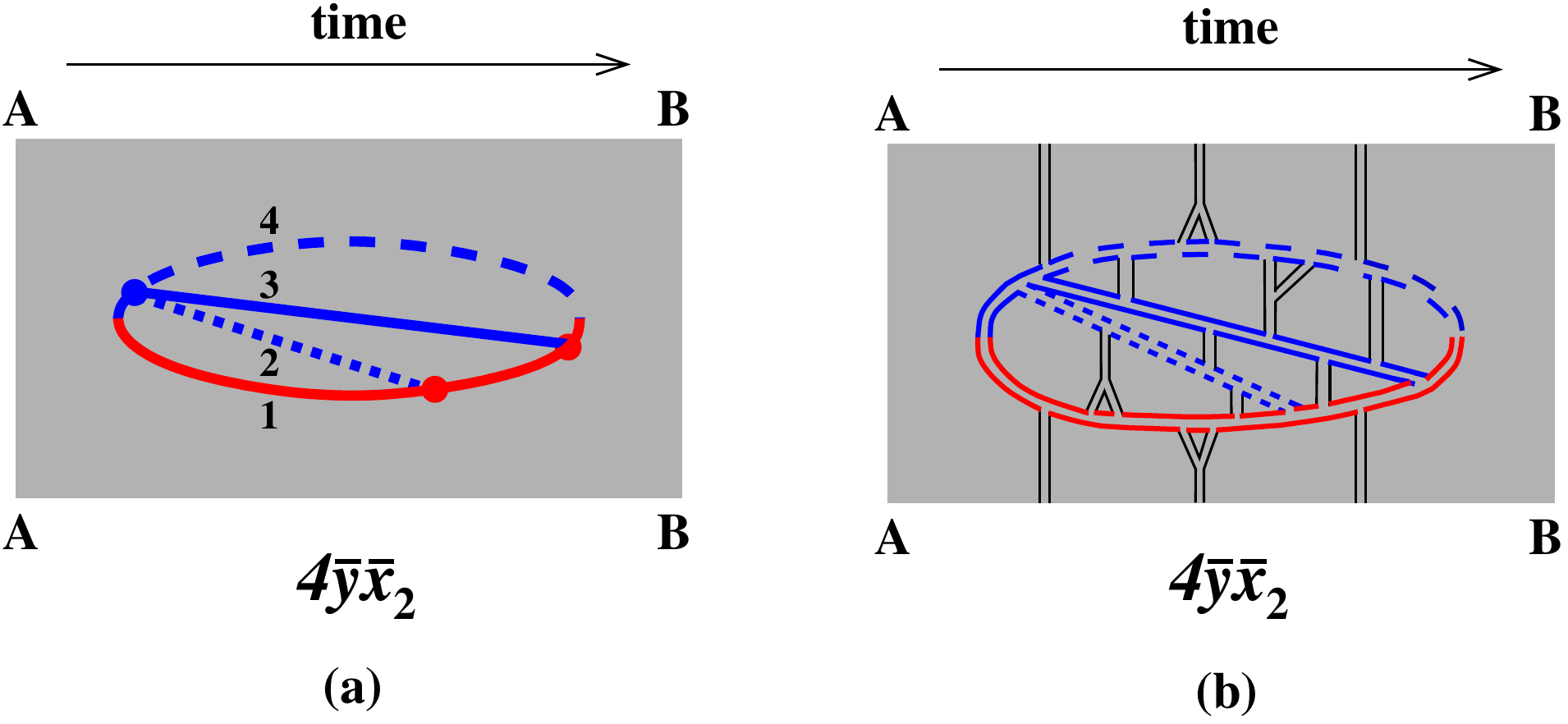}
  \caption{
     \label{fig:4yx2}
     As fig.\ \ref{fig:4yx1} but showing the other distinct
     color routing of $4\bar y\bar x$.
     In our notation, this interference contribution could be referred
     to as either $4\bar y\bar x_2$ or $4\bar y\bar z_1$.
  }
\end {center}
\end {figure}

The diagram for $4\bar y\bar x$ shown in fig.\ \ref{fig:4yx} is
technically symmetric under the permutation
$x \leftrightarrow z$, where $z \equiv 1{-}x{-}y$.
However, in this paper we will work in the large-$\Nc$ limit in order
to simplify the color dynamics of 4-particle evolution.
In this limit, there are two distinct color routings of the
$4\bar y\bar x$ diagram which are not individually $y \leftrightarrow z$
symmetric, just like the situation for the $xy\bar x\bar y$ diagram
discussed in ref.\ \cite{seq}.  We show these two large-$\Nc$ color
routings in figs.\ \ref{fig:4yx1} and \ref{fig:4yx2}, which we will
refer to as $4\bar y\bar x_1$ and $4\bar y\bar x_2$ respectively.
Note that the two routings are related by $x \leftrightarrow z$,
and so we could also call them $4\bar y\bar z_2$ and $4\bar y\bar z_1$
respectively.

Like the situation for the $xy\bar x\bar y$ diagram
discussed in ref.\ \cite{seq}, the distinguishing difference
between the calculation of the two color routings is the
assignment of the longitudinal momentum fractions $x_i$ for the
4-particle part of the evolution, which occurs here for
$t_\four < t < t_\ybx$.  Going around the cylinder depicted in
fig.\ \ref{fig:4yx1}, the first routing $4\bar y\bar x_1$ has
\begin {equation}
  (x_1,x_2,x_3,x_4) = (-1,y,x,1{-}x{-}y) ,
\end {equation}
whereas the second routing $4\bar y\bar x_2$ of fig.\ \ref{fig:4yx2}
has
\begin {equation}
   (x_1,x_2,x_3,x_4) = (-1,y,1{-}x{-}y,x)
   \equiv (\hat x_1,\hat x_2,\hat x_3,\hat x_4) .
\label {eq:xcanonical}
\end {equation}
Note that the ordering of the $x_i$ does not matter until we take the
large-$\Nc$ limit and decide that the 4-particle propagator
$\langle\C_{34}^\ybx,\C_{12}^\ybx,t_\ybx|\C_{34}^\four,\C_{12}^\four,t_\four\rangle$
will henceforth represent only a single color routing.  That is why
the $x_i$ assignment of fig.\ \ref{fig:4yx}, {\it before} we discussed
large-$\Nc$, could represent the entire contribution of $4\bar y\bar
x$, but in our convention after we implement the large-$\Nc$ limit for
discussion of the 4-particle propagator, the same assignment
(\ref{eq:xcanonical}) now represents only a single color routing
(fig.\ \ref{fig:4yx2}).

We will focus on the second routing (\ref{eq:xcanonical})
just because the assignment
$x_i = \hat x_i$ is identical to the
one used for the canonical diagram analyzed in ref.\ \cite{2brem}.
We can obtain the other routing via $x \leftrightarrow z$:
\begin {equation}
   \left[\frac{dI}{dx\,dy}\right]_{4\bar y\bar x}^{\rm total}
   =
   \left[\frac{dI}{dx\,dy}\right]_{4\bar y\bar x_2}
   + [x \leftrightarrow z] .
\label {eq:4routing}
\end {equation}

The details of extracting what pieces of the color and helicity factors
given by fig.\ \ref{fig:dH4point} correspond to which of the two
large-$\Nc$ color routings are a bit untidy.  One can either (i) figure
out how to split up the factors in fig.\ \ref{fig:dH4point} or else (ii)
switch to large-$\Nc$ Feynman rules.  Here we'll take the first option,
as we found it the least confusing way to keep track of overall
normalization factors.

If we label
the gluon lines as $(\ix,\xx,\yx,\zx)$ for the initial, $x$, $y$, and
$z$ bosons, then the color and helicity factors
given by fig.\ \ref{fig:dH4point}
for the 4-point vertex are
\begin {align}
      f^{a_\ix a_\xx e} & f^{a_\yx a_\zx e}
      (\delta_{h_\ix,h_\yx} \delta_{h_\zx,-h_\xx}
        - \delta_{h_\ix,h_\zx} \delta_{h_\xx,-h_\yx})
\nonumber\\ &
    + f^{a_\ix a_\yx e} f^{a_\xx a_\zx e}
      (\delta_{h_\ix,h_\xx} \delta_{h_\yx,-h_\zx}
        - \delta_{h_\ix,h_\zx} \delta_{h_\xx,-h_\yx})
\nonumber\\ &
    + f^{a_\ix a_\zx e} f^{a_\xx a_\yx e}
      (\delta_{h_\ix,h_\xx} \delta_{h_\yx,-h_\zx}
        - \delta_{h_\ix,h_\yx} \delta_{h_\zx,-h_\xx})
    .
\label {eq:color0}
\end {align}
The large-$\Nc$ routing $4\bar y\bar x_2$ of fig.\ \ref{fig:4yx2} corresponds
to  the first term above plus half of the second term,
\begin {align}
      f^{a_\ix a_\xx e} & f^{a_\yx a_\zx e}
      (\delta_{h_\ix,h_\yx} \delta_{h_\zx,-h_\xx}
        - \delta_{h_\ix,h_\zx} \delta_{h_\xx,-h_\yx})
\nonumber\\ &
    + \tfrac12 f^{a_\ix a_\yx e} f^{a_\xx a_\zx e}
      (\delta_{h_\ix,h_\xx} \delta_{h_\yx,-h_\zx}
        - \delta_{h_\ix,h_\zx} \delta_{h_\xx,-h_\yx}) ,
\label {eq:color1}
\end {align}
while the rest of (\ref{eq:color0}) corresponds to the routing
of fig.\ \ref{fig:4yx1}.
The advantage of the large-$\Nc$ limit is that it then allows
us to do a naive color contraction of the vertices in
fig.\ \ref{fig:4yx1}a and \ref{fig:4yx2}a for each routing.%
\footnote{
  A similar use of naive color contractions in large $\Nc$ was made in the
  analysis of ref.\ \cite{2brem} to get eq.\ (4.16) of that reference.
  The various factors of $\Nc$ associated with each additional
  loop caused by an interaction with the medium in figs.\ \ref{fig:4yx1}b
  and \ref{fig:4yx2}b are accounted for in the value of
  the medium parameter $\hat q$.
}
In fig.\ \ref{fig:4yx2}a, (\ref{eq:color1}) is contracted
with adjoint color factors
\begin {equation}
   (T_{\rm A}^{a_\yx})_{a_\ix \bar a} (T_{\rm A}^{a_\xx})_{\bar a a_\zx}
   = - f^{a_\yx a_\ix \bar a} f^{a_\xx \bar a a_\zx}
\label {eq:4yx2color}
\end {equation}
associated with the two 3-point vertices and averaged over
initial color $a_\ix$, giving
\begin {equation}
   - \tfrac12 \CA^2
   (   \delta_{h_\ix,h_\xx} \delta_{h_\yx,-h_\zx}
     + \delta_{h_\ix,h_\yx} \delta_{h_\zx,-h_\xx}
     -2\delta_{h_\ix,h_\zx} \delta_{h_\xx,-h_\yx} )
\label {eq:colorhelicity}
\end {equation}
overall.

Using the rules for 3-gluon vertices, the general expression
(\ref{eq:4yxstart}) then becomes
\begin {align}
   \left[\frac{dI}{dx\,dy}\right]_{4\bar y\bar x_2}
   &=
   - \left( \frac{E}{2\pi} \right)^2
   \int_{t_\four < t_\ybx < t_\xbx}
   \sum_{h_\xx,h_\yx,h_\zx,\bar h}
   \int_{\B^\ybx}
\nonumber\\ &\times
   \tfrac{i}{2} \CA^2 g^4
   (   \delta_{h_\ix,h_\xx} \delta_{h_\yx,-h_\zx}
     + \delta_{h_\ix,h_\yx} \delta_{h_\zx,-h_\xx}
     -2\delta_{h_\ix,h_\zx} \delta_{h_\xx,-h_\yx} )
\nonumber\\ &\times
   \tfrac12 E^{-3/2} \,
   \bcalP_{{-}h_\zx,\bar h,{-}h_\xx}({-}\hat x_3,\hat x_3{+}\hat x_4,{-}\hat x_4)
       \cdot \grad_{\B^\xbx}
   \langle\B^\xbx,t_\xbx|\B^\ybx,t_\ybx\rangle
   \Bigr|_{\B^\xbx=0}
\nonumber\\ &\times
   \tfrac12 E^{-3/2} |\hat x_3+\hat x_4|^{-1} \,
   \bcalP_{{-}\bar h,h_\ix,{-}h_\yx}
               (\hat x_1{+}\hat x_2,{-}\hat x_1,{-}\hat x_2)
          \cdot \grad_{\C_{12}^\ybx}
\nonumber\\ &\qquad\qquad
  \langle\C_{34}^\ybx,\C_{12}^\ybx,t_\ybx|\C_{34}^\four,\C_{12}^\four,t_\four\rangle
   \Bigr|_{\C_{12}^\ybx=0=\C_{34}^\four=\C_{12}^\four; ~ \C_{34}^\ybx=\B^\ybx}
\nonumber\\ &\times
  (2E)^{-2} |\hat x_1\hat x_2\hat x_3 \hat x_4|^{-1/2} |\hat x_3+\hat x_4|^{-1}
  .
\label {eq:4yx2start}
\end {align}
for the routing $4\bar y\bar x_2$.
(See appendix \ref{app:details} for details on the overall sign.)


\subsection {Helicity Sums}
\label {sec:helicity4yx}

For the helicity sums, we need
\begin {multline}
   \sum_{h_\xx,h_\yx,h_\zx,\bar h}
   {\cal P}^{\bar n}_{{-}h_\zx,\bar h,{-}h_\xx}\bigl({-}(1{-}x{-}y),1{-}y,{-}x\bigr) \,
   {\cal P}^{\bar m}_{{-}\bar h,h_\ix,{-}h_\yx}\bigl({-}(1{-}y),1,{-}y\bigr)
\\ \times
   (   \delta_{h_\ix,h_\xx} \delta_{h_\yx,-h_\zx}
     + \delta_{h_\ix,h_\yx} \delta_{h_\zx,-h_\xx}
     -2\delta_{h_\ix,h_\zx} \delta_{h_\xx,-h_\yx} )
   |\hat x_1\hat x_2\hat x_3 \hat x_4|^{-1/2}
\label {eq:calPsum}
\end {multline}
which is equivalent to
\begin {multline}
   \sum_{h_\xx,h_\yx,h_\zx}
   \Bigl[
   \sum_{\bar h}
   {\cal P}^{\bar n}_{\bar h \to h_\zx,h_\xx}\bigl(1{-}y \to 1{-}x{-}y,x\bigr) \,
   {\cal P}^{\bar m}_{h_\ix \to \bar h, h_\yx}\bigl(1 \to 1{-}y,y\bigr)
   \Bigr]^*
\\ \times
   (   \delta_{h_\ix,h_\xx} \delta_{h_\yx,-h_\zx}
     + \delta_{h_\ix,h_\yx} \delta_{h_\zx,-h_\xx}
     -2\delta_{h_\ix,h_\zx} \delta_{h_\xx,-h_\yx} )
   |\hat x_1\hat x_2\hat x_3 \hat x_4|^{-1/2}
   .
\label {eq:calPsum2}
\end {multline}
Note that we have found it convenient to include the
$|\hat x_1\hat x_2\hat x_3 \hat x_4|^{-1/2}$ factor from
(\ref{eq:4yx2start}) here.

By transverse parity invariance, we may average over the initial
helicity.  By transverse rotational invariance, the initial helicity
average of (\ref{eq:calPsum2}) must be of the form
\begin {equation}
   \zeta(x,y) \, \delta^{\bar n\bar m}
\label {eq:zetadelta}
\end {equation}
for some function $\zeta(x,y)$.  Taking the formulas for the
splitting functions ${\cal P}$ from ref.\ \cite{2brem},%
\footnote{
   AI (4.35)
}
we find
\begin {equation}
   \zeta
   = \frac{
        2 x^2 - z^2 - (1{-}y)^4 + 2 y^2 z^2 - x^2 y^2
     }{x^2 y^2 z^2 (1{-}y)^3}
   ,
\label {eq:zeta}
\end {equation}
where $z \equiv 1{-}x{-}y$.
Replacing (\ref{eq:calPsum}) by (\ref{eq:zetadelta})
in (\ref{eq:4yx2start}) gives
\begin {multline}
   \left[\frac{dI}{dx\,dy}\right]_{4\bar y\bar x_2}
   =
   - i \, \frac{\CA^2\alphas^2}{8 E^3} \,
   \frac{\zeta}{|\hat x_3 + \hat x_4|^2}
   \int_{t_\four < t_\ybx < t_\xbx}
   \int_{\B^\ybx}
   \grad_{\B^\xbx}
   \langle\B^\xbx,t_\xbx|\B^\ybx,t_\ybx\rangle
   \Bigr|_{\B^\xbx=0}
\\
   \cdot \grad_{\C_{12}^\ybx}
   \langle\C_{34}^\ybx,\C_{12}^\ybx,t_\ybx|\C_{34}^\four,\C_{12}^\four,t_\four\rangle
   \Bigr|_{\C_{12}^\ybx=0=\C_{34}^\four=\C_{12}^\four; ~ \C_{34}^\ybx=\B^\ybx}
  .
\label {eq:4yx2a}
\end {multline}


\subsection {Harmonic Oscillator Approximation}

Now take the harmonic oscillator approximation.
As reviewed in ref.\ \cite{2brem}, for 3-particle evolution this
corresponds to treating $\langle \B,t|\B',t' \rangle$ as evolution
of a two-dimensional harmonic oscillator with a certain effective
mass $M$ and complex natural frequency $\Omega$.  In the
case of the final 3-particle evolution $t_\ybx < t < t_\xbx$ in
figs.\ \ref{fig:4yx} and \ref{fig:4yx2}, these are \cite{2brem}%
\footnote{
  AI (5.4)
}
\begin {subequations}
\label {eq:MOmf}
\begin {equation}
   M_\fx = \hat x_3 \hat x_4 (\hat x_3{+}\hat x_4) E
   = x(1{-}y)(1{-}x{-}y) E
\label {eq:Mf}
\end {equation}
and
\begin {equation}
   \Omega_\fx
   = \sqrt{ 
     -\frac{i \hat q_{\rm A}}{2E}
     \left( - \frac{1}{\hat x_3 + \hat x_4}
            + \frac{1}{\hat x_4} + \frac{1}{\hat x_3}
            \right)
   }
   = \sqrt{ 
     -\frac{i \hat q_{\rm A}}{2E}
     \left( - \frac{1}{1{-}y} + \frac{1}{x} + \frac{1}{1{-}x{-}y}
            \right)
   } .
\end {equation}
\end {subequations}
Using a harmonic oscillator propagator gives%
\footnote{
  AI (5.9b)
}
\begin {equation}
   \int_{t_\ybx}^{+\infty} dt_\xbx \>
   \grad_{\B^\xbx} \langle\B^\xbx,t_\xbx|\B^\ybx,t_\ybx\rangle
   \biggr|_{\B^\xbx=0}
   =
   - \frac{i M_\fx \B^\ybx}{\pi (B^\ybx)^2} \,
   \exp\bigl(
      - \tfrac12 |M_\fx| \Omega_\fx (B^\ybx)^2
   \bigr) ,
\label {eq:t1int}
\end {equation}
which recasts (\ref{eq:4yx2a}) as
\begin {multline}
   \left[\frac{d\Gamma}{dx\,dy}\right]_{4\bar y\bar x_2}
   =
   - \frac{\CA^2\alphas^2 M_\fx}{8 \pi E^3} \,
   \frac{\zeta}{|\hat x_3 + \hat x_4|^2}
   \int_0^\infty d(\Delta t)
   \int_{\B^\ybx}
   \exp\bigl( - \tfrac12 |M_\fx| \Omega_\fx (B^\ybx)^2 \bigr)
\\ \times
   \frac{\B^\ybx}{(B^\ybx)^2}
   \cdot \grad_{\C_{12}^\ybx}
   \langle\C_{34}^\ybx,\C_{12}^\ybx,\Delta t|\C_{34}^\four,\C_{12}^\four,0\rangle
   \Bigr|_{\C_{12}^\ybx=0=\C_{34}^\four=\C_{12}^\four; ~ \C_{34}^\ybx=\B^\ybx}
  ,
\label {eq:4yx2}
\end {multline}
where $\Delta t \equiv t_\ybx-t_\four$.
We now treat the 4-particle propagator
$\langle\C_{34}^\ybx,\C_{12}^\ybx,\Delta t|\C_{34}^\four,\C_{12}^\four,0\rangle$
just as in section V.C of
ref.\ \cite{2brem}, except that here we have chosen to use the same basis
$(\C_{34},\C_{12})$ in both the bra and the ket.
The propagator is given by
\begin {align}
   \exp\bigl( - \tfrac12 |M_\fx| & \Omega_\fx (C_{34}^\ybx)^2 \bigr) \,
   \langle\C_{34}^\ybx,\C_{12}^\ybx,\Delta t|\C_{34}^\four,\C_{12}^\four,0\rangle
   =
\nonumber\\ &
   (2\pi i)^{-2}
     ({-}x_1 x_2 x_3 x_4)
     |x_3{+}x_4|^2 E^2
   \Omega_+\Omega_- \csc(\Omega_+\Delta t) \csc(\Omega_-\Delta t)
\nonumber\\ & \times
   \exp\Biggl[
     - \frac12
     \begin{pmatrix} \C^\four_{34} \\ \C^\four_{12} \end{pmatrix}^\top \!
       \begin{pmatrix} X_\four & Y_\four \\ Y_\four & Z_\four \end{pmatrix}
       \begin{pmatrix} \C^\four_{34} \\ \C^\four_{12} \end{pmatrix}
     -
     \frac12
     \begin{pmatrix} \C^\ybx_{34} \\ \C^\ybx_{12} \end{pmatrix}^\top \!
       \begin{pmatrix} X_\ybx & Y_\ybx \\ Y_\ybx & Z_\ybx \end{pmatrix}
       \begin{pmatrix} \C^\ybx_{34} \\ \C^\ybx_{12} \end{pmatrix}
\nonumber\\ & \qquad
     +
     \begin{pmatrix} \C^\four_{34} \\ \C^\four_{12} \end{pmatrix}^\top \!
       \begin{pmatrix} X_{\four\ybx} & Y_{\four\ybx} \\
                       \Ybar_{\four\ybx} & Z_{\four\ybx} \end{pmatrix}
       \begin{pmatrix} \C^\ybx_{34} \\ \C^\ybx_{12} \end{pmatrix}
   \Biggr],
\label {eq:Cprop4}
\end {align}
where we have included on the left-hand side of (\ref{eq:Cprop4})
the additional factor
$\exp\bigl( - \tfrac12 |M_\fx| \Omega_\fx (B^\ybx)^2 \bigr)
 = \exp\bigl( - \tfrac12 |M_\fx| \Omega_\fx (C_{34}^\ybx)^2 \bigr)$
from
(\ref{eq:4yx2}) because that makes the definitions of the symbols
$X$, $Y$, and $Z$ more convenient for later use.
Those symbols are then given by
\begin {subequations}
\label {eq:4XYZdef}
\begin {align}
   \begin{pmatrix} X_\four & Y_\four \\ Y_\four & Z_\four \end{pmatrix}
   &\equiv
     - i a_\four^{-1\top} \uOmega \cot(\uOmega\,\Delta t)\, a_\four^{-1} ,
\\
   \begin{pmatrix} X_\Ax & Y_\Ax \\ Y_\Ax & Z_\Ax \end{pmatrix}
   &\equiv
   \begin{pmatrix} |M_\fx|\Omega_\fx & 0 \\ 0 & 0 \end{pmatrix}
     - i a_\ybx^{-1\top} \uOmega \cot(\uOmega\,\Delta t)\, a_\ybx^{-1} ,
\\
   \begin{pmatrix} X_{\four\Ax} & Y_{\four\Ax} \\
                   \Ybar_{\four\Ax} & Z_{\four\Ax} \end{pmatrix}
   &\equiv
   - i a_\four^{-1\top} \uOmega \csc(\uOmega\,\Delta t) \, a_\ybx^{-1} ,
\end {align}
\end {subequations}
where (given our choice of basis at the 4-point vertex)
\begin {equation}
   a_\four = a_\ybx =
   \begin{pmatrix} C^+_{34} & C^-_{34} \\ C^+_{12} & C^-_{12} \end{pmatrix} .
\label {eq:4adef}
\end {equation}
Above,
$\uOmega \equiv
 \bigl(
 \begin{smallmatrix} \Omega_+ & \\ & \Omega_- \end{smallmatrix}
 \bigr)$.
Formulas from \cite{2brem}
for the two 4-particle evolution frequencies $\Omega_\pm$ and
the corresponding normal modes $(C_{34}^\pm,C_{12}^\pm)$ are
collected in Appendix \ref{app:modes}.

Using (\ref{eq:Cprop4}) in (\ref{eq:4yx2}) gives
\begin {multline}
   \left[\frac{d\Gamma}{dx\,dy}\right]_{4\bar y\bar x_2}
   =
   \frac{\CA^2\alphas^2 M_\fx}{8 \pi E^3} \,
   \frac{\zeta}{|\hat x_3 + \hat x_4|^2}
   \int_0^\infty d(\Delta t)
   \int_{\B^\ybx}
   (2\pi i)^{-2} (-\hat x_1 \hat x_2 \hat x_3 \hat x_4)
   |\hat x_3 + \hat x_4|^2 E^2
\\ \times
   \Omega_+ \Omega_- \csc(\Omega_+\,\Delta t) \csc(\Omega_-\,\Delta t)
   Y_\ybx \exp\bigl( - \tfrac12 X_\ybx (B^\ybx)^2 \bigr) .
\end {multline}
The Gaussian $\B^\ybx$ integral is straightforward, yielding
\begin {equation}
   \left[\frac{d\Gamma}{dx\,dy}\right]_{4\bar y\bar x_2}
   =
   - \frac{\CA^2\alphas^2 M_\fx}{16 \pi^2 E} \,
   (-\hat x_1 \hat x_2 \hat x_3 \hat x_4)
   \zeta
   \int_0^\infty d(\Delta t) \>
   \Omega_+ \Omega_- \csc(\Omega_+\,\Delta t) \csc(\Omega_-\,\Delta t)
   \frac{Y_\ybx}{X_\ybx} \,.
\label {eq:4yx2result}
\end {equation}
Our final result for the $4\bar y\bar x$ diagram is the above
formula together with the corresponding version of (\ref{eq:4routing}),
\begin {equation}
   \left[\frac{d\Gamma}{dx\,dy}\right]_{4\bar y\bar x}^{\rm total}
   =
   \left[\frac{d\Gamma}{dx\,dy}\right]_{4\bar y\bar x_2}
   + [x \leftrightarrow z] .
\end {equation}


\section {The Other Diagrams}
\label {sec:other}

\subsection {The \boldmath$\bar y\bar x4$ diagram}

The $\bar y\bar x 4$ diagram is the third diagram of
fig.\ \ref{fig:4subset2}.  Instead of going through an explicit calculation,
we can relate the answer for this diagram to the $4\bar y\bar x$ diagram
computed in the last section, along the lines of how the
$x\bar y y\bar x$ and $xy\bar y\bar x$ diagrams of fig.\ \ref{fig:subset2}
were related in ref.\ \cite{2brem}.

\begin {figure}[t]
\begin {center}
 \includegraphics[scale=0.6]{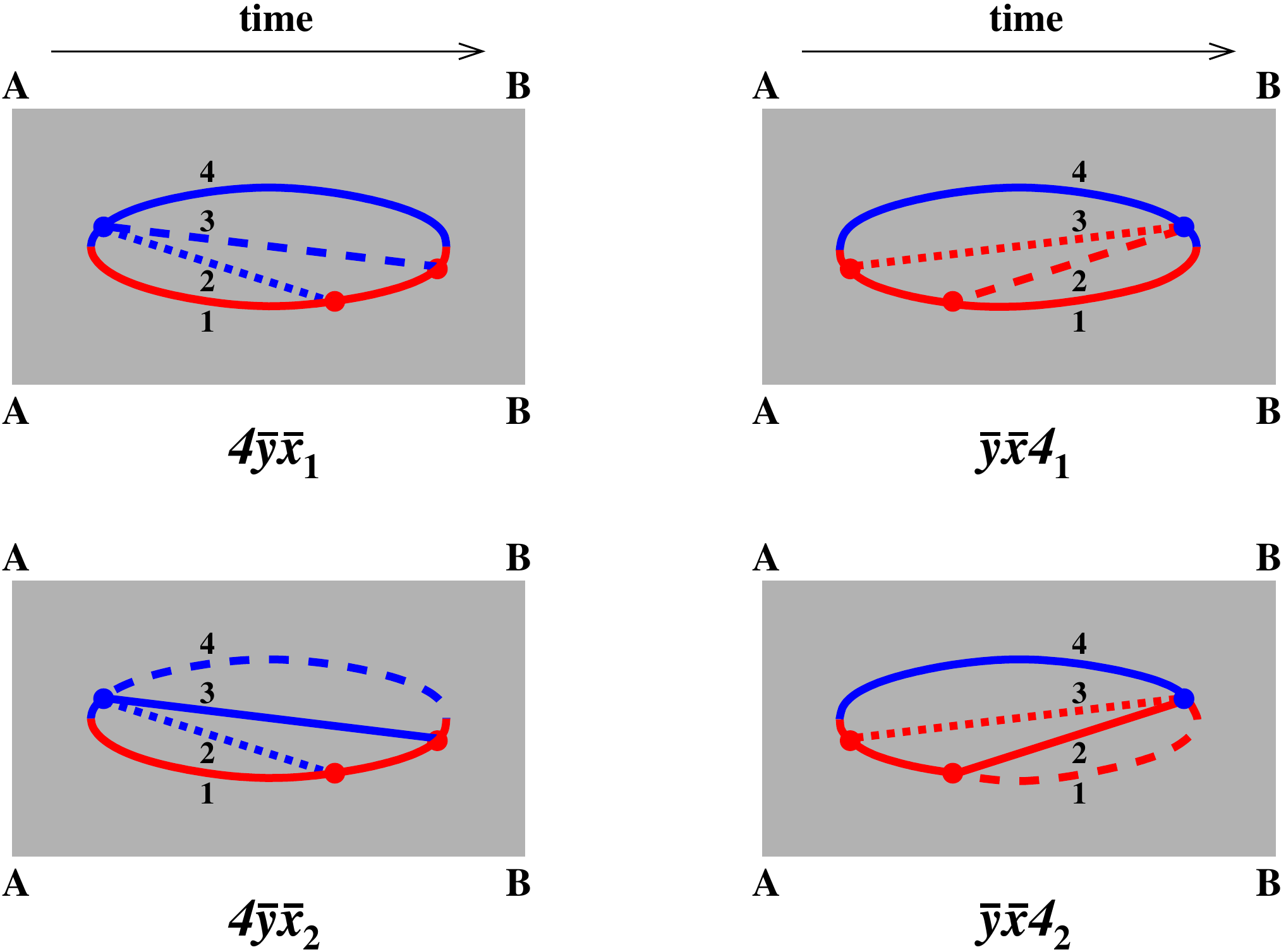}
  \caption{
     \label{fig:mirror}
     The two color routings of $4\bar y\bar x$ (left) compared to those
     of $\bar y\bar x4$ (right).  There are many topologically-equivalent
     ways to draw the same diagram: we've chosen to draw the $\bar y\bar x4$
     diagrams above in a way that gives a straightforward pictorial
     correspondence to our rule (\ref{eq:xmirror})
     for going from the $4\bar y\bar x$ diagrams on the left to the
     $\bar y\bar x4$ diagrams on the right.
  }
\end {center}
\end {figure}

The first thing to note is that all three diagrams shown explicitly in
fig.\ \ref{fig:4subset2} have the same factors of helicity contractions and
DGLAP splitting functions associated with their vertices---these factors
are unaffected by the time ordering of the 4-point vertex in the amplitude
relative to the two vertices in the conjugate amplitude.
As to the rest of the computation, 
note that the diagrams
$\bar y\bar x 4$ and $4\bar x\bar y$ in fig.\ \ref{fig:4subset2}
look like mirror images of each other except for the identification
of which gluon has which momentum fraction.
For each color routing, we show one way of making this change
of identification in fig.\ \ref{fig:mirror}.  There, when reflecting
$4\bar y\bar x$ into $\bar y\bar x4$, we change
\begin {equation}
   (x_1,x_2,x_3,x_4) = (-1,y,x,1{-}x{-}y)
\end {equation}
to
\begin {equation}
   (x_1,x_2,x_3,x_4) = \bigl(-(1{-}x{-}y),-x,-y,1\bigr)
\end {equation}
for the first color routing,
and
\begin {equation}
   (x_1,x_2,x_3,x_4) = (-1,y,1{-}x{-}y,x)
\end {equation}
to
\begin {equation}
   (x_1,x_2,x_3,x_4) = \bigl(-x,-(1{-}x{-}y),-y,1\bigr)
\label {eq:xmirror2}
\end {equation}
for the second.  Both cases can be summarized as
\begin {equation}
   (x_1,x_2,x_3,x_4) \to (-x_4,-x_3,-x_2,-x_1) .
\label {eq:xmirror}
\end {equation}
We also need to appropriately change the mass $M$ used for the
3-particle part of the evolution.  As for similar diagram transformations
in ref.\ \cite{2brem}, this will be taken care of automatically if we
write this mass in terms of the 4-particle $x_i$ as in (\ref{eq:Mf}):
\begin {equation}
  M = x_3 x_4 (x_3{+}x_4) E ,
\label {eq:Mmirror}
\end {equation}
which, for example, gives $M=x(1{-}y)(1{-}x{-}y)E$ (\ref{eq:Mf}) for
3-particle evolution in the $4\bar y\bar x_2$ case of
$(x_1,x_2,x_3,x_4) = (\hat x_1,\hat x_2,\hat x_3,\hat x_4)$
and gives $M= -y(1{-}y)E$ for the corresponding
$\bar y\bar x4_2$ case
$(x_1,x_2,x_3,x_4) = (-\hat x_4,-\hat x_3,-\hat x_2,-\hat x_1)$.

The upshot is that we can convert the result for $4\bar y\bar x$ into
a result for $\bar y\bar x 4$ by (i) making the change (\ref{eq:xmirror}) to
the 4-particle $x_i$, (ii) always using the form (\ref{eq:Mmirror})
for the 3-particle evolution mass,
and (iii) leaving $\zeta(x,y)$ unchanged.%
\footnote{
  Because we only care about the real part of interference diagrams,
  the negation of the $x_i$ in (\ref{eq:xmirror}) does not matter at
  the end of the day.  Negation of all the $x_i$ simply has the effect of
  complex conjugation of the diagram (i.e. swapping
  the amplitude and conjugate amplitude).
}
For the sake of readers wary of the glibness of the above argument,
we give a more straightforward derivation
of $\bar y \bar x 4_2$ in appendix \ref{app:yx4} and verify that the
result is the same.


\subsection {The \boldmath$\bar y4\bar x$ diagram}

Now consider the $\bar y4\bar x$ interference contribution, depicted
by the second diagram in fig.\ \ref{fig:4subset2}.  The starting point,
analogous to (\ref{eq:4yxstart}), is
\begin {multline}
   \left[\frac{dI}{dx\,dy}\right]_{\bar y4\bar x}
   =
   \left( \frac{E}{2\pi} \right)^2
   \int_{t_\ybx < t_\four < t_\xbx}
   \sum_{\rm pol.}
   \langle|i\,\overline{\delta H}|\B^\xbx\rangle \,
   \langle\B^\xbx,t_\xbx|\B^\four,t_\four\rangle
\\ \times
   \langle\B^\four|i\,\overline{\delta H}|\B'{}^\four\rangle \,
   \langle\B'{}^\four,t_\four|\B^\ybx,t_\ybx\rangle
   \langle\B^\ybx|i\,\overline{\delta H}|\rangle
   .
\label {eq:y4xstart}
\end {multline}
We will not need to work out the explicit normalization of the
4-gluon vertex matrix element
$\langle\B^\four|i\,\overline{\delta H}|\B'{}^\four\rangle$
(though we give it in Appendix \ref{app:4matrix}) because we
will find that (\ref{eq:y4xstart}) is zero.
The important point is that the helicity factors and splitting
factors ${\cal P}$ are the same as they were for $4\bar y\bar x$
in section \ref{sec:helicity4yx}, and so, using fig.\ \ref{fig:dH},
\begin {multline}
   \left[\frac{dI}{dx\,dy}\right]_{\bar y4\bar x}
   \propto
   \zeta \delta^{\bar m\bar n}
   \int_{t_\ybx < t_\four < t_\xbx}
   \sum_{\rm pol.}
   \nabla^{\bar n}_{\B^\xbx}
   \langle\B^\xbx,t_\xbx|\B^\four,t_\four\rangle
      \Bigl|_{\B^\xbx = 0 = \B^\four}
\\ \times
   \nabla^{\bar m}_{\B^\ybx}
   \langle\B'{}^\four,t_\four|\B^\ybx,t_\ybx\rangle
      \Bigl|_{\B^\ybx = 0 = \B'{}^\four}
   .
\label {eq:y4x}
\end {multline}
The reason that $\B^\four$ and $\B'{}^\four$ are set to zero above is because
in 3-particle evolution (analogous to the earlier statement about
4-particle evolution), the transverse positions $\b_i$ in our problem
are implicitly constrained by the condition
$x_1 \b_1 + x_2 \b_2 + x_3 \b_3 = 0$ with $x_1 + x_2 + x_3 = 0$.
(See section III of ref.\ \cite{2brem}.)  One may use this constraint
to show that there is but one relevant transverse degree of freedom for
the three transverse positions in 3-particle evolution:%
\footnote{
  AI (2.29)
}
\begin {equation}
   \B \equiv \frac{\b_1-\b_2}{(x_1+x_2)}
   = \frac{\b_2-\b_3}{(x_2+x_3)}
   = \frac{\b_3-\b_1}{(x_3+x_1)} \,.
\label {eq:B}
\end {equation}
So, in our application, when any two of the three particles are coincident,
then $\B = 0$ and all three of the particles are necessarily coincident.

But now we can see the result.  The factors
\begin {equation}
   \nabla^{\bar n}_{\B^\xbx}
   \langle\B^\xbx,t_\xbx|\B^\four,t_\four\rangle
      \Bigl|_{\B^\xbx = 0 = \B^\four}
   \qquad \mbox{and} \qquad
   \nabla^{\bar m}_{\B^\ybx}
   \langle\B'{}^\four,t_\four|\B^\ybx,t_\ybx\rangle
      \Bigl|_{\B^\ybx = 0 = \B'{}^\four}
\end {equation}
must both be zero by parity, and so the $\bar y4\bar x$ contribution
(\ref{eq:y4x}) vanishes.


\subsection {The \boldmath$4\bar 4$ diagram}

\begin {figure}[t]
\begin {center}
  \includegraphics[scale=0.55]{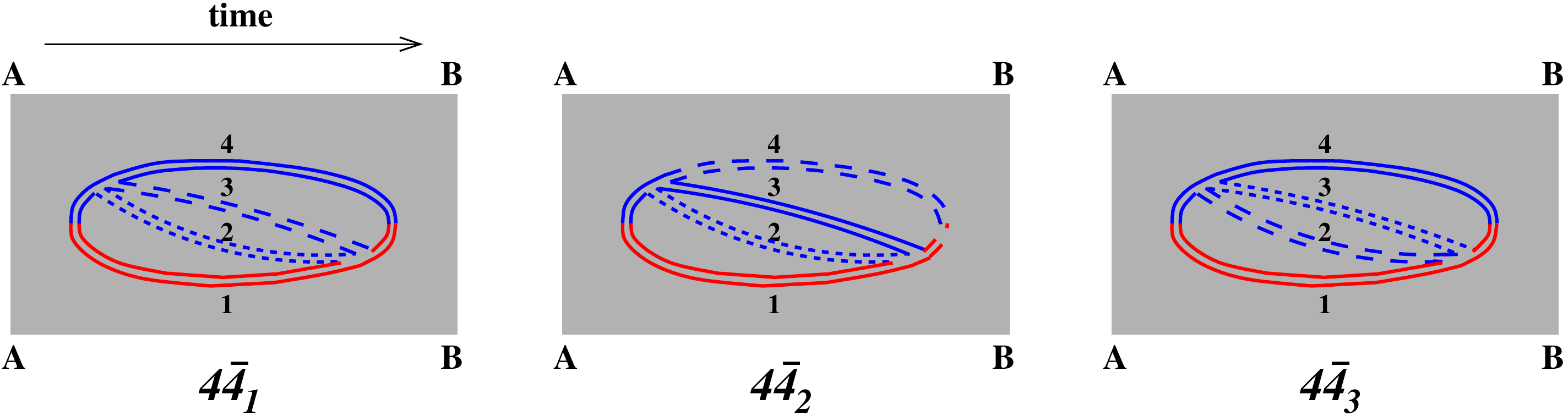}
  \caption{
     \label{fig:44color}
     The three distinct large-$\Nc$ color routings of the $4\bar 4$ interference
     diagram, drawn on a cylinder in large-$\Nc$ double-line notation.
     Other possible ways to draw color
     routings are equivalent.  As usual, long-dashed and short-dashed lines
     refer to the gluons with momentum fraction $x$ and $y$ respectively.
  }
\end {center}
\end {figure}

The $4\bar 4$ diagram, shown in fig.\ \ref{fig:44subset2},
is formally given by
\begin {multline}
   \left[\frac{dI}{dx\,dy}\right]_{4\bar4}
   =
   \left( \frac{E}{2\pi} \right)^2
   \int_{t_\four < t_\fourb}
   \sum_{\rm pol.}
   \langle|i\,\overline{\delta H}|\C_{34}^\fourb,\C_{12}^\fourb\rangle \,
   \langle\C_{34}^\fourb,\C_{12}^\fourb,t_\fourb|\C_{34}^\four,\C_{12}^\four,t_\four\rangle
\\ \times
   \langle\C_{34}^\four,\C_{12}^\four|{-}i\,\delta H|\rangle .
\label {eq:44start}
\end {multline}
This diagram has
three distinct large-$\Nc$ color routings, shown in fig.\ \ref{fig:44color},
which are related by permutations of the
three final-state gluons $(x,y,1{-}x{-}y)$.

The helicity and color factors associated with the 4-gluon matrix elements
do not depend on the longitudinal momentum fractions (e.g.\ $x$ and $y$)
of the various gluons and so, when summed over polarizations and colors,
give the exact same helicity/color factor for each of the three color
routings of fig.\ \ref{fig:44color}.  Each is therefore a third of the
total helicity/color factor $S$ we would get in a vacuum calculation,
where we would not need to split the calculation into different color
routings but could simply square and initial-state
average the color/helicity factors (\ref{eq:color0})
of the 4-point vertex:
\begin {align}
   S & \equiv \frac{1}{2\dA} \sum_{\mbox{\tiny$h$'s}} \sum_{\rm color} \Bigl[
      f^{a_\ix a_\xx e} f^{a_\yx a_\zx e}
      (\delta_{h_\ix,h_\yx} \delta_{h_\zx,-h_\xx}
        - \delta_{h_\ix,h_\zx} \delta_{h_\xx,-h_\yx})
\nonumber\\ & \qquad\qquad\qquad
    + f^{a_\ix a_\yx e} f^{a_\xx a_\zx e}
      (\delta_{h_\ix,h_\xx} \delta_{h_\yx,-h_\zx}
        - \delta_{h_\ix,h_\zx} \delta_{h_\xx,-h_\yx})
\nonumber\\ & \qquad\qquad\qquad
    + f^{a_\ix a_\zx e} f^{a_\xx a_\yx e}
      (\delta_{h_\ix,h_\xx} \delta_{h_\yx,-h_\zx}
        - \delta_{h_\ix,h_\yx} \delta_{h_\zx,-h_\xx})
    \Bigr]^2
\nonumber\\
   & = 9 \CA^2
\end {align}
(where $\dA$ is the dimension of the adjoint representation).
So each color routing has a corresponding factor of
$S/3 = 3 \CA^2$.

We will focus on the second color routing $4\bar 4_2$,
which is convenient because it again
corresponds to our canonical choice (\ref{eq:xcanonical}),
\begin {equation}
   (x_1,x_2,x_3,x_4) = (-1,y,1{-}x{-}y,x)
   \equiv (\hat x_1,\hat x_2,\hat x_3,\hat x_4) .
\end {equation}
The corresponding contribution to (\ref{eq:44start}) is
\begin {multline}
   \left[\frac{dI}{dx\,dy}\right]_{4\bar 4_2}
   =
   \left( \frac{E}{2\pi} \right)^2
   \int_{t_\four < t_\fourb}
   3\CA^2 g^4 (2 E)^{-4}|\hat x_1 \hat x_2 \hat x_3 \hat x_4|^{-1}
        |\hat x_3 + \hat x_4|^{-2}
\\ \times
   \langle\C_{34}^\fourb,\C_{12}^\fourb,t_\fourb|\C_{34}^\four,\C_{12}^\four,t_\four\rangle
   \Bigr|_{{\rm all}~ \C_{ij} = 0}
  .
\end {multline}
From (\ref{eq:Cprop4}),
\begin {multline}
   \langle\C_{34}^\fourb,\C_{12}^\fourb,\Delta t|\C_{34}^\four,\C_{12}^\four,0\rangle
   \Bigr|_{{\rm all}~ \C_{ij} = 0}
   =
\\
   (2\pi i)^{-2}
     ({-}x_1 x_2 x_3 x_4)
     |x_3{+}x_4|^2 E^2
   \Omega_+\Omega_- \csc(\Omega_+\Delta t) \csc(\Omega_-\Delta t)
   ,
\end {multline}
and so
\begin {equation}
   \left[\frac{d\Gamma}{dx\,dy}\right]_{4\bar 4_2}
   =
   - \frac{3\CA^2 \alphas^2}{16 \pi^2}
   \int_0^\infty d(\Delta t) \>
   \Omega_+\Omega_- \csc(\Omega_+\,\Delta t) \csc(\Omega_-\,\Delta t)
  .
\label {eq:44}
\end {equation}
We may then sum all the color routings by adding appropriate permutations:
\begin {equation}
   \left[\frac{d\Gamma}{dx\,dy}\right]_{4\bar4}^{\rm total}
   =
   \left[\frac{d\Gamma}{dx\,dy}\right]_{4\bar4_2}
   + [x \leftrightarrow z]
   + [y \leftrightarrow z] .
\end {equation}

Note that $\left[\frac{d\Gamma}{dx\,dy}\right]_{4\bar 4}$ should be
positive since it is the medium average of the
magnitude-squared of something (the amplitude for double
bremsstrahlung via the 4-gluon vertex in the background of the medium).
The numerical result shown in fig.\ \ref{fig:4pointResult44} verifies
this is the case.%
\footnote{
  One might think of checking that the {\it total} double bremsstrahlung
  rate $d\Gamma/dx\,dy$, which is also the medium average of the magnitude
  squared of something (the total amplitude for double bremsstrahlung),
  is also positive.  However, as discussed in ref.\ \cite{seq}, the
  total $d\Gamma/dx\,dy$ is formally infinite in our calculation, and
  the physically relevant quantity is instead $\Delta\,d\Gamma/dx\,dy$
  defined by (\ref{eq:DeltaDef}).  The latter is a {\it difference}\/ of
  two positive quantities and so can have either sign (as seen in
  fig.\ \ref{fig:4pointResult}).
}

We also note in passing that we can evaluate (\ref{eq:44}) analytically
in the limit that one of the final-state gluons in soft.
For $y \ll x$ and $z$, the result for the total contribution
of fig.\ \ref{fig:44subset2} (i.e.\ adding in the conjugate diagrams) is
\begin {equation}
   \left[\frac{d\Gamma}{dx\,dy}\right]_{(44)}
   \equiv
   2\Re \left[\frac{d\Gamma}{dx\,dy}\right]_{4\bar 4}
   \simeq
   6\Re \left[\frac{d\Gamma}{dx\,dy}\right]_{4\bar 4_2}
   \simeq
   \frac{9\CA^2 \alphas^2 \ln2}{16 \pi^2}
   \sqrt{\frac{\hat q_{\rm A}}{yE}}
   \qquad
   (y\ll x,z)
   .
\label {eq:44limit}
\end {equation}
(See appendix \ref{app:details}.)


\section {Summary of Formula}
\label {sec:summary}

The total result for the correction $\Delta\,d\Gamma/dx\,dy$ due to
overlapping formation times is
\begin {equation}
  \Delta\frac{d\Gamma}{dx\,dy} =
  \left[ \frac{d\Gamma}{dx\,dy} \right]_{\rm crossed}
  + \left[ \Delta\frac{d\Gamma}{dx\,dy} \right]_{\rm seq}
  + \left[ \frac{d\Gamma}{dx\,dy} \right]_\four
  + \left[ \frac{d\Gamma}{dx\,dy} \right]_\ff ,
\label {eq:total}
\end {equation}
where $[d\Gamma/dx\,dy]_{\rm crossed}$ and $[\Delta\,d\Gamma/dx\,dy]_\seq$
are given respectively in ref.\ \cite{2brem,dimreg}
and ref.\ \cite{seq}.  For completeness, we have summarized those
formulas in Appendix \ref{app:crossseq}.
The contributions new to this paper, involving one or more
4-gluon vertices, are summarized below.


\subsection {Diagrams with one 4-gluon vertex}

The diagrams of fig.\ \ref{fig:4subset2} (including all permutations,
large-$\Nc$ color routings, and conjugates) give the following
contribution to $d\Gamma/dx\>dy$:
\begin {align}
   \left[ \frac{d\Gamma}{dx\>dy} \right]_\four
   = \quad
   & {\cal A}_\four(x,y) + {\cal A}_\four(1{-}x{-}y,y)
                        + {\cal A}_\four(x,1{-}x{-}y)
\nonumber\\
   + ~ &
   {\cal A}_\four(y,x) + {\cal A}_\four(y,1{-}x{-}y)
                      + {\cal A}_\four(1{-}x{-}y,x)
\label {eq:dGamma4}
\end {align}
where ${\cal A}_\four(x,y)$ is
the result for one color routing of
$4\bar y\bar x + \bar y 4\bar x + \bar y\bar x 4$ plus
conjugates.
We will write this as
\begin {equation}
   {\cal A}_\four(x,y)
   \equiv
   \int_0^{+\infty} d(\Delta t) \>
        2 \Re \bigl( B_\four(x,y,\Delta t) \bigr)
\end {equation}
where
\begin {align}
   B_\four(x,y,\Delta t) &=
       D_\four(\hat x_1,\hat x_2,\hat x_3,\hat x_4,\zeta,\Delta t)
       + D_\four(-\hat x_4,-\hat x_3,-\hat x_2,-\hat x_1,\zeta,\Delta t)
\nonumber\\
   &=
       D_\four({-}1,y,1{-}x{-}y,x,\zeta,\Delta t)
       + D_\four(-x,-(1{-}x{-}y),-y,1,\zeta,\Delta t)
\label {eq:B4}
\end {align}
corresponds to
(i) the $4\bar y\bar x_2$ color routing of
$4\bar y\bar x$ plus
(ii) the related color routing $\bar y\bar x4_2$ of $\bar y\bar x4$.
$\zeta=\zeta(x,y)$ is given by (\ref{eq:zeta}).
Each of the terms in (\ref{eq:B4}) is given by
\begin {align}
   D_\four(x_1,&x_2,x_3,x_4,\zeta,\Delta t) =
\nonumber\\ &
   - \frac{\CA^2\alphas^2 M_\fx}{16 \pi^2 E} \,
   (-x_1 x_2 x_3 x_4)
   \zeta
   \Omega_+ \Omega_- \csc(\Omega_+\,\Delta t) \csc(\Omega_-\,\Delta t)
   \frac{Y_\ybx}{X_\ybx} \,,
\label {eq:D4}
\end {align}
which is the integrand of (\ref{eq:4yx2result}).
Here, the $X,Y,Z$ are defined by (\ref{eq:4XYZdef}) and
(\ref{eq:4adef}), with
\begin {align}
   M_\fx &= x_3 x_4 (x_3{+}x_4) E ,
\\
   \Omega_\fx
   &= \sqrt{ 
     -\frac{i \hat q_{\rm A}}{2E}
     \left( \frac{1}{x_3} + \frac{1}{x_4}
            - \frac{1}{x_3{+}x_4}
     \right)
   } .
\end {align}
As mentioned earlier, explicit formulas for
the 4-particle evolution frequencies
$\Omega_\pm$ in terms of $(x_1,x_2,x_3,x_4)$ are collected in
Appendix \ref{app:modes}.

Unlike for the crossed and sequential diagrams analyzed in
refs.\ \cite{2brem,seq}, it is unnecessary to explicitly subtract the
vacuum contribution from $D_\four$.  That's because the vacuum limit
$\hat q \to 0$ (and so $\Omega_\pm \to 0$ and $\Omega_\fx \to 0$) of
(\ref{eq:D4}) already vanishes.

Also unlike the crossed and sequential diagrams \cite{2brem,seq}, there are no
$1/\Delta t$ divergences associated with the individual
diagrams of fig.\ \ref{fig:4subset2},
and so there are no
``pole'' term contributions that need to be included in
${\cal A}_\four$ above.


\subsection {Diagrams with two 4-gluon vertices}

The diagrams of fig.\ \ref{fig:44subset2} give the following
contribution:
\begin {equation}
   \left[ \frac{d\Gamma}{dx\>dy} \right]_\ff
   =
   {\cal A}_\ff(x,y) + {\cal A}_\ff(1{-}x{-}y,y)
                        + {\cal A}_\ff(x,1{-}x{-}y)
\label {eq:dGamma44}
\end {equation}
where ${\cal A}_\ff(x,y)$ is
the result for one color routing of
$4\bar 4$ plus conjugate.
We write this as
\begin {equation}
   {\cal A}_\ff(x,y)
   \equiv
   \int_0^{+\infty} d(\Delta t) \>
        2 \Re \bigl( B_\ff(x,y,\Delta t) \bigr)
\end {equation}
where
\begin {align}
   B_\ff(x,y,\Delta t) &=
       C_\ff(\hat x_1,\hat x_2,\hat x_3,\hat x_4,\Delta t)
   =
       C_\ff({-}1,y,1{-}x{-}y,x,\Delta t)
\end {align}
corresponds to the color routing $4\bar 4_2$
with vacuum subtraction.
The vacuum subtraction is
\begin {equation}
   C_\ff = D_\ff - \lim_{\hat q\to 0} D_\ff ,
\end {equation}
where $D_\ff$ is the unsubtracted result extracted from
(\ref{eq:44}),
\begin {equation}
   D_\ff(x_1,x_2,x_3,x_4,\Delta t) =
   - \frac{3\CA^2 \alphas^2}{16 \pi^2} \,
   \Omega_+\Omega_- \csc(\Omega_+\,\Delta t) \csc(\Omega_-\,\Delta t)
   .
\end {equation}

Again, there are no $1/\Delta t$ divergences associated with the
diagrams here, and so there are no
``pole'' term contributions that need to be included in
${\cal A}_\ffb$ above.


\section {Conclusion}
\label {sec:conclusion}

We have now completed the calculation of the overlapping formation
time correction to double bremsstrahlung for the process $g \to ggg$
of emitting two real bremsstrahlung gluons from an initial gluon.
The size of interference terms involving 4-gluon vertices had to be
computed (i) for completeness and (ii) to see how big they are.
However, the conclusion we can take from the numerical results of
figs.\ \ref{fig:4pointResult}--\ref{fig:4pointResult44} is that their
effect on the result is small and one would not go far wrong in
ignoring them, at least insofar as $\Delta\,d\Gamma/dx\,dy$ is concerned.

An important reason for calculating the overlapping formation
time correction is to test whether it is large or small for
realistic value of $\alphas$.  It is already known that the
corrections due to soft bremsstrahlung ($y \ll 1$) are large due
to large logarithms but that such soft corrections can be resummed
into a running value of $\hat q$ that depends on energy
\cite{Wu0,Blaizot1,Blaizot,Iancu,Wu,Iancu2}.
But what about the contribution from overlapping hard bremsstrahlung,
which cannot be absorbed into $\hat q$?
In the thick-medium approximation used here, these corrections
are controlled by the value of $\alphas$ at scales of order%
\footnote{
  See, for example, the comments in section I.E of ref.\ \cite{2brem}.
}
$Q_\perp \sim (\hat q E)^{1/4}$.
An answer concerning the size of these non-absorbable corrections
will need to wait longer until we are in a position
to calculate an infrared-safe physical quantity characterizing
shower development, which will
require (i) including the effects of virtual corrections to single
bremsstrahlung and (ii) consistent factorization of the effects of soft
bremsstrahlung into $\hat q$.


\begin{acknowledgments}

This work was supported, in part, by the U.S. Department
of Energy under Grant No.~DE-SC0007984.

\end{acknowledgments}

\appendix

\section {More details on some formulas}
\label {app:details}

\paragraph*{Eq.\ (\ref{eq:4yx2start}):}
The overall sign of this formula arises as follows, similar to the
discussion of AI (4.16) in Appendix A of ref.\ \cite{2brem}.  Consider
first the rule associated with the $t=t_\ybx$ vertex in
fig.\ \ref{fig:4yx} (remembering that the ordering of $x_i$ used
in that figure was chosen to match the ordering of the large-$\Nc$
color routing $4\bar y\bar x_2$ of fig.\ \ref{fig:4yx2}).
According to the rules of fig.\ \ref{fig:dH}, this vertex comes
with a factor of
$(T_R^{a_k})_{a_j a_i} \grad\delta^{(2)}(\bcalB_{ji})$, with
lines $(i,j,k)$ identified as in the figure.  Using the cyclic
permutation identity
$\bcalB_{ji}=\bcalB_{kj}=\bcalB_{ik}$ noted in the caption,
and comparing fig.\ \ref{fig:dH} to the $t=t_\ybx$ vertex in
fig.\ \ref{fig:4yx}, we can identify these factors as
$(T_R^{a_k})_{a_j a_i} \grad\delta^{(2)}(\bcalB_{kj})
 = (T_{\rm A}^{a_\yx})_{a_\ix\bar a} \grad\delta^{(2)}(\C_{21})$.
Similarly, the vertex at $t=t_\xbx$ comes with a factor of
$(T_R^{a_k})_{a_j a_i} \grad\delta^{(2)}(\bcalB_{ik})
 = (T_{\rm A}^{a_\xx})_{\bar a a_\zx} \grad\delta^{(2)}(\C_{34})$.
Since we have identified $\C_{34}$ with $\B$ in (\ref{eq:4yx2start}),
the latter is
$(T_{\rm A}^{a_\xx})_{\bar a a_\zx} \grad\delta^{(2)}(\B)$.
The color factors
$(T_{\rm A}^{a_\yx})_{a_\ix\bar a} (T_{\rm A}^{a_\xx})_{\bar a a_\zx}$
from these two vertices (and the signs that
arise from them) have already been accounted
for in (\ref{eq:4yx2color}), which has already been combined
with the 4-gluon vertex factor (and its signs)
in (\ref{eq:colorhelicity}).  We are left with
the $\delta$-function factors 
$\grad\delta^{(2)}(\C_{21})\, \grad\delta^{(2)}(\B)$.
Since $\C_{21} = -\C_{12}$, these may be rewritten as
\begin {equation}
  - \grad\delta^{(2)}(\C_{12})\, \grad\delta^{(2)}(\B) ,
\label {eq:sign}
\end {equation}
which is the form used in (\ref{eq:4yx2start}), where both
$\C_{12}$ and $\B$ have been integrated by parts.
This minus sign combines with the minus sign in (\ref{eq:colorhelicity})
and the $\mp = -$ in fig.\ \ref{fig:dH4point} to give
the overall minus sign in (\ref{eq:4yx2start}).

\paragraph*{Eq.\ (\ref{eq:44limit}):}
In the limit that $y$ is small compared to both $x$ and $z\equiv 1{-}x{-}y$,
the formulas for the 4-particle frequencies $\Omega_\pm$ collected in
appendix \ref{app:modes} satisfy, for the case $x_i = \hat x_i$,
\begin {equation}
   \Omega_- \ll \Omega_+ \simeq
   \Omega_y \equiv \sqrt{- \frac{i\hat q_{\rm A}}{2 y E}} \,.
\end {equation}
The factor of $\csc(\Omega_+ \,\Delta t)$ in (\ref{eq:44}) means that
the integrand is negligible unless $\Omega_+ \Delta t \lesssim 1$, in which
case $\Omega_- \Delta t \ll 1$.  So the integral may be approximated as
\begin {equation}
   \left[\frac{d\Gamma}{dx\,dy}\right]_{4\bar 4_2}
   \simeq
   - \frac{3\CA^2 \alphas^2}{16 \pi^2}
   \int_0^\infty \frac{d(\Delta t)}{\Delta t} \>
   \Omega_y \csc(\Omega_y\,\Delta t)
  .
\end {equation}
This approximation is the same for all three color routings.
Correspondingly multiplying by 3, and then adding in the conjugate
diagram $\bar44$ by taking twice the real part,
\begin {equation}
   \left[\frac{d\Gamma}{dx\,dy}\right]_\ff
   \simeq
   - \frac{9\CA^2 \alphas^2}{8 \pi^2}
   \Re \int_0^\infty \frac{d(\Delta t)}{\Delta t} \>
   \Omega_y \csc(\Omega_y\,\Delta t)
  .
\end {equation}
As we do with all diagrams, we now subtract out the vacuum contribution
$\hat q\to 0$ (i.e.\ $\Omega_y \to 0$), leaving
\begin {align}
   \left[\frac{d\Gamma}{dx\,dy}\right]_\ff
   &\simeq
   - \frac{9\CA^2 \alphas^2}{8 \pi^2}
   \Re
   \int_0^\infty \frac{d(\Delta t)}{\Delta t}
   \left[ \Omega_y \csc(\Omega_y\,\Delta t) - \frac{1}{\Delta t} \right]
\nonumber\\
   &=
   - \frac{9\CA^2 \alphas^2}{8 \pi^2}
   \Re\left( i \Omega_y
   \int_0^\infty \frac{d\tau}{\tau}
   \left[ \frac{1}{\sh\tau} - \frac{1}{\tau} \right]
   \right)
\nonumber\\
   &=
   \frac{9\CA^2 \alphas^2 \ln2}{8 \pi^2}
   \Re(i \Omega_y) ,
\end {align}
which gives (\ref{eq:44limit}).


\section {The 4-gluon matrix element}
\label {app:4matrix}

\subsection{ \boldmath$\langle \C_{34},\C_{12} | \delta H |\rangle$ }
\label {app:CCH}

\begin {figure}[t]
\begin {center}
  \includegraphics[scale=0.6]{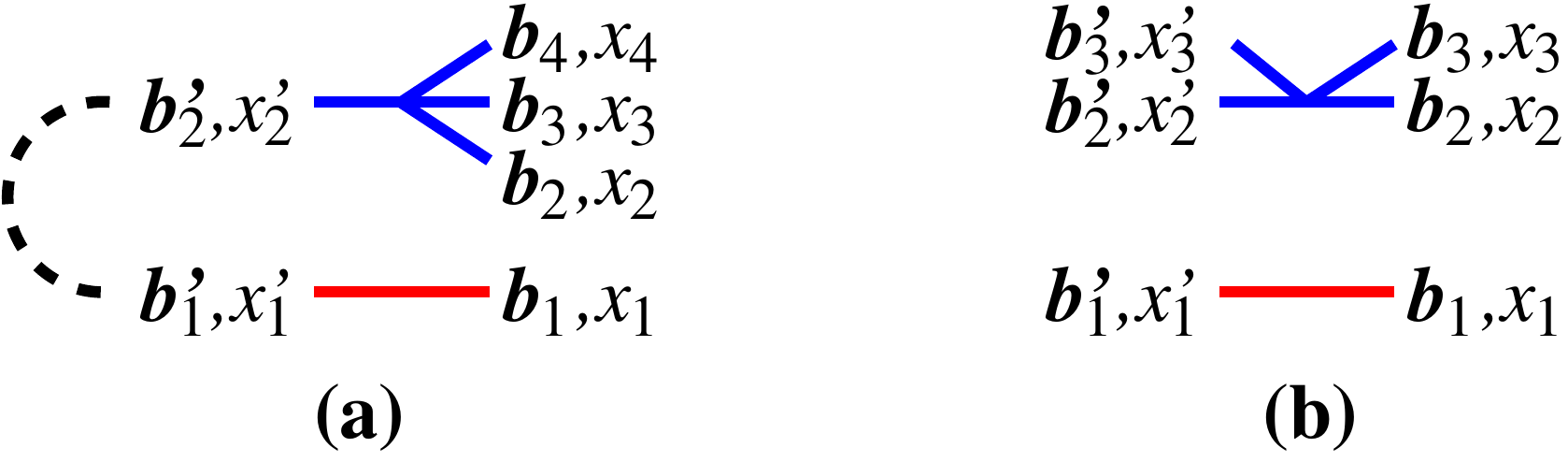}
  \caption{
     \label{fig:dH4labels}
     The notation used in (a) appendix \ref{app:CCH} and (b) appendix
     \ref{app:BHB} to label different particles' states immediately
     before and after a 4-gluon vertex.  The dashed connection in
     (a) indicates the fact that in this case the initial particles
     in the amplitude and conjugate amplitude represent the same particle
     (and so, for instance, $\b_1'=\b_2'$ and, given our conventions,
     $x_1' = - x_2'$).
  }
\end {center}
\end {figure}

To derive the matrix element $\langle \C_{34},\C_{12} | \delta H |\rangle$,
we will follow the method used for deriving other matrix elements in
Appendix B of ref.\ \cite{2brem}.
We start in a description of states where we individually
distinguish each high-energy particle, using the conventions of
fig.\ \ref{fig:dH4labels}a.
First, the $\delta H$ matrix element in the amplitude,
written conventionally in terms of the individual particles
in the Hilbert space $\Hilbert$ (as opposed to the Hilbert space
$\bar\Hilbert\otimes\Hilbert$ used to simultaneously describe particles
in the amplitude and conjugate amplitude), is
\begin {equation}
  \langle \b_2,\b_3,\b_4 | \delta H | \b_2' \rangle
  = {\mathfrak H} \,
    \delta^{(2)}(\b_2{-}\b_2') \, \delta^{(2)}(\b_3{-}\b_2) \,
    \delta^{(2)}(\b_4{-}\b_2)
\label {eq:bbb}
\end {equation}
with
\begin {align}
  {\mathfrak H} \equiv
       g^2 & \Bigl[
           f^{a_1 a_2 e} f^{a_3 a_4 e}
             (\delta_{h_1,-h_3} \delta_{h_4,-h_2} - \delta_{h_1,-h_4} \delta_{h_2,-h_3})
\nonumber\\ & \quad
         + f^{a_1 a_3 e} f^{a_2 a_4 e}
             (\delta_{h_1,-h_2} \delta_{h_3,-h_4} - \delta_{h_1,-h_4} \delta_{h_2,-h_3})
\nonumber\\ & \quad
         + f^{a_1 a_4 e} f^{a_2 a_3 e}
             (\delta_{h_1,-h_2} \delta_{h_3,-h_4} - \delta_{h_1,-h_3} \delta_{h_4,-h_2})
       \Bigr]
\nonumber\\ &
       \times
       (2|x_1|E)^{-1/2} (2|x_2|E)^{-1/2} (2|x_3|E)^{-1/2} (2|x_4|E)^{-1/2}
   .
\label {eq:frakH}
\end {align}
(\ref{eq:bbb}) is the usual relativistic formula except for a few small
differences.  The factors of $(2 E_i)^{-1/2} = (2|x_i|E)^{-1/2}$ for
each particle above are included because we use non-relativistic rather
than relativistic normalization for the states.  We have written the
rule in transverse $\b$-space instead of transverse momentum space, so
there are $\delta$-functions requiring the points to be coincident at
the vertex instead of a $\delta$-function for overall transverse
momentum conservation.  We have assumed that the longitudinal
momenta have already been chosen to satisfy longitudinal momentum
conservation, and we have (just as in ref.\ \cite{2brem}) chosen a
normalization of our states where we implicitly drop the corresponding
momentum-space $\delta(p_{2z}'{-}p_{2z}{-}p_{3z}{-}p_{4z})$.
Finally, we have used the fact that the initial state represents
a single on-shell particle to link the color and helicity of
particle $2'$ to that of $1'$ and thus, via fig.\ \ref{fig:dH4labels}a,
to particle $1$.  We have accordingly chosen to label
the corresponding color and
helicity indices in (\ref{eq:frakH}) by $1$ instead of by $2'$.
The convention used for the flow of helicity here is that of
fig.\ \ref{fig:dH4point}.
The $\delta_{\cdots}\delta_{\cdots}$
terms in (\ref{eq:frakH}) for helicity come from
contracting the usual factors $g_{\mu\nu} g_{\alpha\beta}$ in the Feynman rule
for the 4-point vertex with normalized helicity polarizations
$\epsilon_{(h)}^\mu$ for each particle.

The corresponding matrix element in the space
$\bar\Hilbert\otimes\Hilbert$ that includes the particle in the
conjugate amplitude is
\begin {equation}
  \langle \b_1,\b_2,\b_3,\b_4 | \delta H | \b_1',\b_2' \rangle
  =
  \langle \b_2,\b_3,\b_4 | \delta H | \b_2' \rangle \,
  \delta^{(2)}(\b_1{-}\b_1') .
\label {eq:bbbb1}
\end {equation}
Next we want to use the symmetry of the problem to project each state
onto a subspace with two fewer degrees of freedom,
as discussed in
AI section III and AI Appendix B \cite{2brem}.  Using the
notation of that reference,
\begin {equation}
  \langle \{\C_{ij}\} | \delta H | \rangle
  =
  \frac{1}{\tildeV_\perp} \int_{\Delta\b}
  \langle \b_1, \b_2,\b_3,\b_4 | \delta H
       | \b_1'{+}\Delta\b, \b_2'{+}\Delta\b \rangle,
\end {equation}
where it is understood that both the initial and final positions
satisfy the constraint $\sum_i x_i\b_i=0$ and
where $\tildeV_\perp$ is a formally infinite normalization given by
\begin {equation}
   \tildeV_\perp \equiv
   \delta^{(2)}({\textstyle\sum} x_i \b_i) \Bigr|_{\sum x_i\b_i = 0} .
\label {eq:Vtilde}
\end {equation}
Using (\ref{eq:bbb}) and (\ref{eq:bbbb1}),
\begin {align}
  \langle \{\C_{ij}\} | \delta H | \rangle
  &=
  \frac{\mathfrak H}{\tildeV_\perp} \int_{\Delta\b}
  \delta^{(2)}(\b_1-\b_1'{-}\Delta\b) \,
    \delta^{(2)}(\b_2{-}\b_2'{-}\Delta\b) \,
    \delta^{(2)}(\b_3{-}\b_2) \,
    \delta^{(2)}(\b_4{-}\b_2)
\nonumber\\
  &=
  \frac{\mathfrak H}{\tildeV_\perp} \,
  \delta^{(2)}(\b_{12}{-}\b_{12}') \,
    \delta^{(2)}(\b_{32}) \,
    \delta^{(2)}(\b_{42})
  ,
\end {align}
where $\b_{ij} \equiv \b_i-\b_j$.
The initial state $|\b_1',\b_2'\rangle$ satisfies the constraint
$x_1' \b_1' + x_2' \b_2' = 0$ with $x_1'+x_2' = 0$, and therefore
$\b_{12}' = 0$, giving
\begin {equation}
  \langle \{\C_{ij}\} | \delta H | \rangle
  =
  \frac{\mathfrak H}{\tildeV_\perp} \,
  \delta^{(2)}(\b_{12}) \,
    \delta^{(2)}(\b_{32}) \,
    \delta^{(2)}(\b_{42})
  .
\label {eq:CH1}
\end {equation}
Given the presence of the other two $\delta$-functions, the first one
can be rewritten as
\begin {equation}
  \delta^{(2)}(\b_{12})
  = x_1^2 \, \delta^{(2)}( x_1\b_{12} {+} x_3\b_{32} {+} x_4\b_{42} )
  = x_1^2 \, \delta^{(2)}( x_1\b_1 {+} x_2\b_2 {+} x_3\b_3 {+} x_4\b_4 ) ,
\label {eq:deltasub}
\end {equation}
where the last equality uses
\begin {subequations}
\label {eq:constraints}
\begin {equation}
   x_1 + x_2 + x_3 + x_4 = 0 .
\end {equation}
Since
\begin {equation}
   x_1\b_1 + x_2\b_2 + x_3\b_3 + x_4\b_4 = 0
\label {eq:xbconstraint}
\end {equation}
\end {subequations}
as well,
the substitution (\ref{eq:deltasub}) in
(\ref{eq:CH1}) gives
\begin {equation}
  \langle \{\C_{ij}\} | \delta H | \rangle
  =
  \mathfrak H x_1^2 \,
    \delta^{(2)}(\b_{32}) \,
    \delta^{(2)}(\b_{42})
\label{eq:CH2}
\end {equation}
by (\ref{eq:Vtilde}).
Because of the constraints (\ref{eq:constraints}), the variables
$\b_{32}$ and $\b_{42}$ are related to
$\C_{12} \equiv \b_{12}/(x_1{+}x_2)$ and
$\C_{34} \equiv \b_{34}/(x_3{+}x_4)$ by%
\footnote{
  AI (5.14)
}
\begin {subequations}
\begin {align}
  \b_{32} &= -\b_{23} = x_1 \C_{12} + x_4 \C_{34} ,
\\
  \b_{42} &= -\b_{24} = x_1 \C_{12} - x_3 \C_{34} ,
\end {align}
\end {subequations}
and the Jacobean for the change of variables is
$\partial(\b_{32},\b_{42})/\partial(\C_{12},\C_{34}) = [x_1(x_3{+}x_4)]^2.$
So (\ref{eq:CH2}) can be rewritten as
\begin {equation}
  \langle \{\C_{ij}\} | \delta H | \rangle
  =
  \frac{\mathfrak H}{(x_3{+}x_4)^2} \,
    \delta^{(2)}(\C_{12}) \,
    \delta^{(2)}(\C_{34}) .
\end {equation}
Changing normalization as in ref.\ \cite{2brem},%
\footnote{
  AI (4.23)
}
\begin {equation}
   |\C_{34},\C_{12}\rangle \equiv |x_3{+}x_4| \, \bigl| \{\C_{ij}\} \bigr\rangle ,
\end {equation}
then gives the matrix element (\ref{eq:dHCC}) displayed in the main text.


\subsection{ \boldmath$\langle \B | \delta H |\B'\rangle$ }
\label {app:BHB}

\begin {figure}[t]
\begin {center}
  \begin{picture}(415,136)(10,10)
  \put(25,20){\includegraphics[scale=0.7]{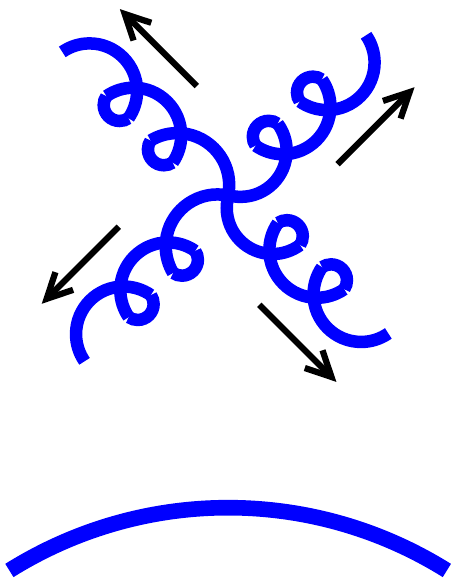}}
  \put(17,55){$\b_i,a_i$}
  \put(102,55){$\b_j,a_j$}
  \put(17,140){$\b_l,a_l$}
  \put(102,140){$\b_k,a_k$}
  \put(30,88){$h_i$}
  \put(77,58){$h_j$}
  \put(107,108){$h_k$}
  \put(57,135){$h_l$}
  \put(17,30){$\b_m$}
  \put(115,30){$\b_n$}
  \put(145,75){$\displaystyle{
     = \quad
     \begin {aligned}
       \mp i g^2 & \Bigl[
           f^{a_i a_j e} f^{a_k a_l e}
             (\delta_{h_i,-h_k} \delta_{h_l,-h_j} - \delta_{h_i,-h_l} \delta_{h_j,-h_k})
     \\ & \quad
         + f^{a_i a_k e} f^{a_j a_l e}
             (\delta_{h_i,-h_j} \delta_{h_k,-h_l} - \delta_{h_i,-h_l} \delta_{h_j,-h_k})
     \\ & \quad
         + f^{a_i a_l e} f^{a_j a_k e}
             (\delta_{h_i,-h_j} \delta_{h_k,-h_l} - \delta_{h_i,-h_k} \delta_{h_l,-h_j})
       \Bigr]
     \\ &
       \times
       (2E)^{-2} |x_i x_j x_k x_l|^{-1/2} |x_m|^{-2} \,
       \delta^{(2)}(\bcalB_{ij}) \, \delta^{(2)}(\bcalB_{kl})
     \end {aligned}
  }$}
  \end{picture}
  \caption{
     \label{fig:dH4point2}
     As fig.\ \ref{fig:dH4point} but for the case with an
     an additional spectator
     (e.g.\ as in the $\bar y4\bar x$ diagram of fig.\ \ref{fig:4subset2}).
  }
\end {center}
\end {figure}

We do not need to figure out the correct normalization of
the matrix element $\langle \B | \delta H |\B'\rangle$
for this paper, but we do so here just for the sake of completeness.
The corresponding diagrammatic rule we will find is shown in
fig.\ \ref{fig:dH4point2}.

Analogous to (\ref{eq:bbb}), start with the amplitude matrix element
\begin {equation}
  \langle \b_2,\b_3 | \delta H | \b_2',\b_3' \rangle
  = {\mathfrak H}' \,
    \delta^{(2)}(\b_2{-}\b_2') \, \delta^{(2)}(\b_3{-}\b_3') \,
    \delta^{(2)}(\b_3{-}\b_2) ,
\label {eq:bbbx}
\end {equation}
using the labeling of fig.\ \ref{fig:dH4labels}b.
Here ${\mathfrak H}'$ is the same as (\ref{eq:frakH}) except that
the indices $1$ and $4$ are replaced by $2'$ and $3'$.
Including the particle in the conjugate amplitude,
\begin {equation}
  \langle \b_1,\b_2,\b_3 | \delta H | \b_1',\b_2',\b_3' \rangle
  =
  \langle \b_2,\b_3 | \delta H | \b_2',\b_3' \rangle \,
  \delta^{(2)}(\b_1{-}\b_1') .
\end {equation}
Projecting the number of degrees of freedom in each state from 3
to 1 as in ref.\ \cite{2brem},
\begin {align}
  \langle \B | \delta H |\B'\rangle &=
  \frac{1}{\tildeV_\perp} \int_{\Delta\b}
  \langle \b_1 \b_2,\b_3 | \delta H
       | \b_1'{+}\Delta\b, \b_2'{+}\Delta\b, \b_3'{+}\Delta\b \rangle
\nonumber\\
  &=
  \frac{{\mathfrak H}'}{\tildeV_\perp} \,
  \delta^{(2)}(\b_{21}{-}\b_{21}') \,
    \delta^{(2)}(\b_{31}{-}\b_{31}') \,
    \delta^{(2)}(\b_{32})
  .
\end {align}
Using the constraint $x_1'{+}x_2'{+}x_3' = 0$ and
the primed version of the
relationships (\ref{eq:B}) defining $\B$, 
\begin {align}
  \langle \B | \delta H |\B'\rangle &=
  \frac{{\mathfrak H}'}{\tildeV_\perp} \,
  \delta^{(2)}(\b_{21}{-}x_3'\B') \,
    \delta^{(2)}(\b_{31}{+}x_2'\B') \,
    \delta^{(2)}(x_1'\B')
\nonumber\\
  &=
  \frac{{\mathfrak H}'}{(x_1')^2\tildeV_\perp} \,
  \delta^{(2)}(\b_{21}) \,
    \delta^{(2)}(\b_{31}) \,
    \delta^{(2)}(\B')
  .
\end {align}
Given the other $\delta$-functions, the middle one can be rewritten as
\begin {equation}
  \delta^{(2)}(\b_{31})
  = x_3^2 \, \delta^{(2)}(x_2\b_{21}{+}x_3\b_{31})
  = x_3^2 \, \delta^{(2)}(x_1\b_1{+}x_2\b_2{+}x_3\b_3) .
\end {equation}
From the constraint $x_1\b_1{+}x_2\b_2{+}x_3\b_3{=}0$ and (\ref{eq:Vtilde}),
we then have
\begin {align}
  \langle \B | \delta H |\B'\rangle &=
  \frac{{\mathfrak H}'}{(x_1')^2} \,
  x_3^2 \, \delta^{(2)}(\b_{21}) \,
  \delta^{(2)}(\B')
\nonumber\\
  &=
  \frac{{\mathfrak H}'}{x_1^2} \,
  \delta^{(2)}(\B) \,
  \delta^{(2)}(\B')
  ,
\end {align}
where in the last line we've used $x_3 = -(x_2{+}x_1)$ and have noted
that $x'_1 = x_1$ in the diagram of fig.\ \ref{fig:dH4labels}b.


\section {Relating \boldmath$\bar y\bar x 4$ to \boldmath$4\bar y\bar x$}
\label {app:yx4}

\begin {figure}[t]
\begin {center}
  \begin{picture}(250,150)(0,0)
  \put(0,0){\includegraphics[scale=0.5]{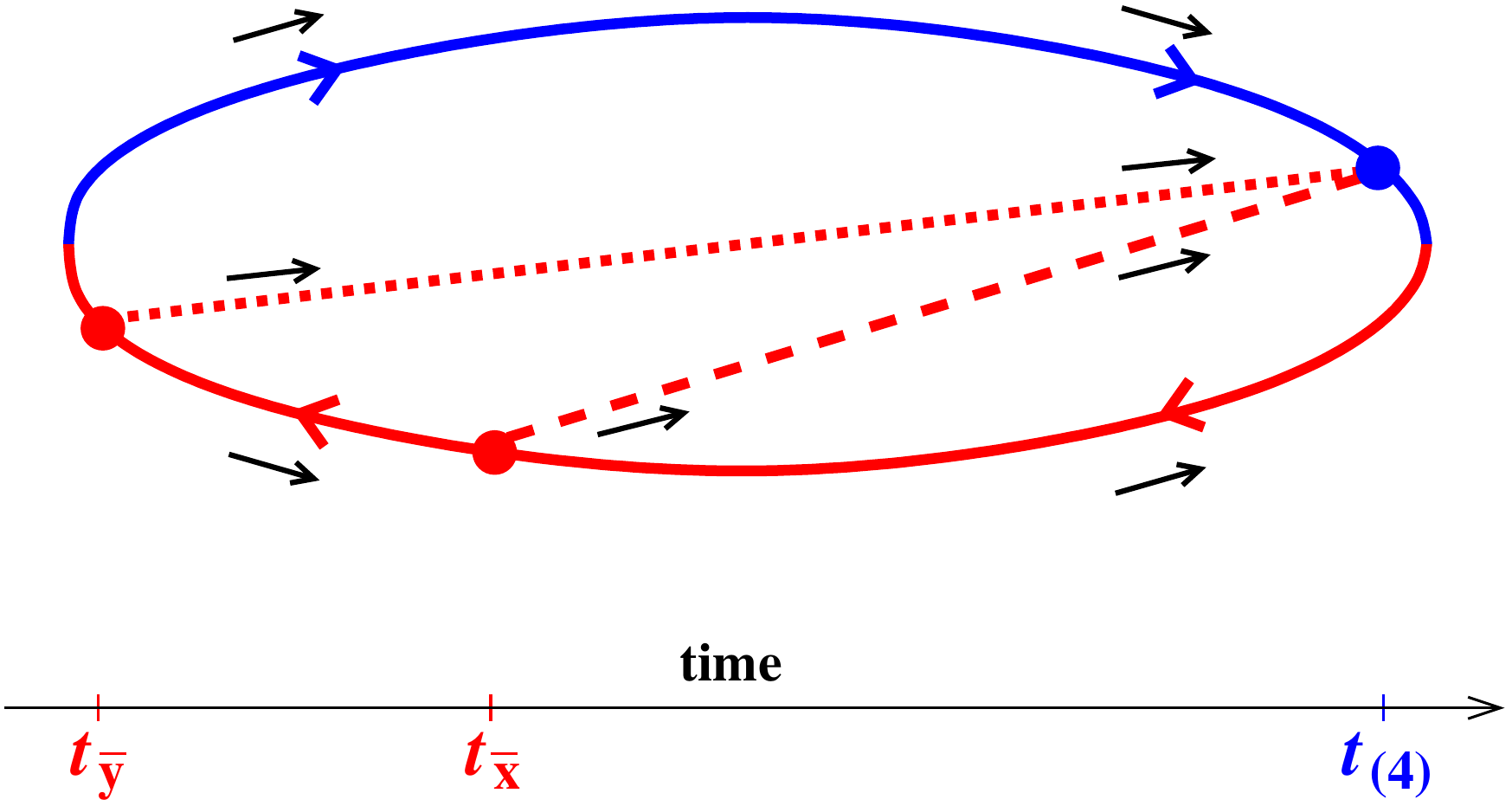}}
  \put(150,44){${-}h_\zx,\,\dot x_2{=}{-}z{=}{-}(1{-}x{-}y)$}
  \put(165,78){${-}h_{\rm x},\,\dot x_1{=}{-}x$}
  \put(125,105){${-}h_{\rm y},\,\dot x_3{=}{-}y$}
  \put(170,138){$h_{\rm i},\,\dot x_4{=}1$}
  \put(-20,44){${-}\bar h,\,{-}(\dot x_3{+}\dot x_4){=}{-}(1{-}y)$}
  \end{picture}
  \caption{
     \label{fig:yx4}
     Labeling conventions for
     the $\bar y\bar x 4$ interference diagram.
  }
\end {center}
\end {figure}
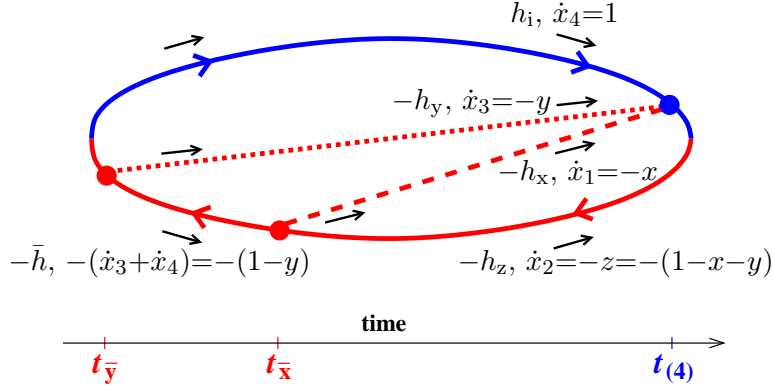

In this appendix, we sketch what happens if we evaluate $\bar y\bar x4$ by
following the same steps used for $4\bar y\bar x$ in section
\ref{sec:4yx}.
Fig.\ \ref{fig:yx4} shows the analog of fig.\ \ref{fig:4yx}.
Here, the $\dot x_i$ are the $x_i$ of (\ref{eq:xmirror2}):
\begin {equation}
   (\dot x_1,\dot x_2,\dot x_3,\dot x_4) \equiv \bigl(-x,-(1{-}x{-}y),-y,1\bigr)
   = (-\hat x_4, -\hat x_3, -\hat x_2, -\hat x_1).
\label {eq:xmirror2b}
\end {equation}

The starting point analogous to (\ref{eq:4yxstart}) is
\begin {align}
   \left[\frac{dI}{dx\,dy}\right]_{\bar y\bar x4}
   =
   \left( \frac{E}{2\pi} \right)^2
   \int_{t_\ybx < t_\xbx < t_\four}
   &
   \sum_{\rm pol.}
   \langle|{-}i\,\delta H|\C_{34}^\four,\C_{12}^\four\rangle
   \langle\C_{34}^\four,\C_{12}^\four,t_\four|\C_{34}^\xbx,\C_{12}^\xbx,t_\xbx\rangle
\nonumber\\ &\times
   \langle\C_{34}^\xbx,\C_{12}^\xbx|i\,\overline{\delta H}|\B^\xbx\rangle \,
   \langle\B^\xbx,t_\xbx|\B^\ybx,t_\ybx\rangle
   \langle\B^\ybx|i\,\overline{\delta H}|\rangle .
\end {align}
Following the same arguments as in section \ref{sec:color}, the
expression for the large-$\Nc$ color routing $\bar y\bar x 4_2$ of
fig.\ \ref{fig:mirror} is
\begin {align}
   \left[\frac{dI}{dx\,dy}\right]_{\bar y\bar x 4_2}
   &=
   - \left( \frac{E}{2\pi} \right)^2
   \int_{t_\ybx < t_\xbx < t_\four}
   \sum_{h_\xx,h_\yx,h_\zx,\bar h}
   \int_{\B^\xbx}
\nonumber\\ &\times
   \tfrac{i}{2} \CA^2 g^4
   (   \delta_{h_\ix,h_\xx} \delta_{h_\yx,-h_\zx}
     + \delta_{h_\ix,h_\yx} \delta_{h_\zx,-h_\xx}
     -2\delta_{h_\ix,h_\zx} \delta_{h_\xx,-h_\yx} )
\nonumber\\ &\times
   \tfrac12 E^{-3/2} |\dot x_3+\dot x_4|^{-1} \,
   \bcalP_{{-}h_\zx,\bar h,{-}h_\xx}(\dot x_2,-\dot x_1{-}\dot x_2,\dot x_1)
          \cdot \grad_{\C_{12}^\xbx}
\nonumber\\ &\qquad\qquad
  \langle\C_{34}^\four,\C_{12}^\four,t_\four|\C_{34}^\xbx,\C_{12}^\xbx,t_\xbx\rangle
   \Bigr|_{\C_{12}^\xbx=0=\C_{34}^\four=\C_{12}^\four; ~ \C_{34}^\xbx=\B^\xbx}
\nonumber\\ &\times
  (2E)^{-2} |\dot x_1\dot x_2\dot x_3 \dot x_4|^{-1/2} |\dot x_3+\dot x_4|^{-1}
\nonumber\\ &\times
   \tfrac12 E^{-3/2} \,
   \bcalP_{{-}\bar h,h_\ix,{-}h_\yx}
               ({-}\dot x_3{-}\dot x_4,\dot x_4,\dot x_3)
       \cdot \grad_{\B^\ybx}
   \langle\B^\xbx,t_\xbx|\B^\ybx,t_\ybx\rangle
   \Bigr|_{\B^\ybx=0}
  ,
\end {align}
analogous to (\ref{eq:4yx2start}).
The helicity sums are exactly the same in terms of $x$ and $y$ as
those in section \ref{sec:helicity4yx}, giving
\begin {multline}
   \left[\frac{dI}{dx\,dy}\right]_{\bar y\bar x 4_2}
   =
   - i \, \frac{\CA^2\alphas^2}{8 E^3} \,
   \frac{\zeta}{|\dot x_3 + \dot x_4|^2}
   \int_{t_\ybx < t_\xbx < t_\four}
   \int_{\B^\xbx}
\\
   \grad_{\C_{12}^\xbx}
   \langle\C_{34}^\four,\C_{12}^\four,t_\four|\C_{34}^\xbx,\C_{12}^\xbx,t_\xbx\rangle
   \Bigr|_{\C_{12}^\xbx=0=\C_{34}^\four=\C_{12}^\four; ~ \C_{34}^\xbx=\B^\xbx}
\\
   \cdot
   \grad_{\B^\ybx}
   \langle\B^\xbx,t_\xbx|\B^\ybx,t_\ybx\rangle
   \Bigr|_{\B^\ybx=0}
\end {multline}
as the analogy to (\ref{eq:4yx2a}).
In the harmonic oscillator approximation,%
\footnote{
  AI (5.9)
}
\begin {equation}
   \int_{-\infty}^{t} dt' \>
   \grad_{\B'} \langle\B,t|\B',t'\rangle
   \biggr|_{\B'=0}
   =
   - \frac{i M \B}{\pi B^2} \,
   \exp\bigl(
      - \tfrac12 |M| \Omega B^2
   \bigr) ,
\end {equation}
and so
\begin {subequations}
\label {eq:yx4}
\begin {multline}
   \left[\frac{d\Gamma}{dx\,dy}\right]_{\bar y\bar x 4_2}
   =
   - \frac{\CA^2\alphas^2 \dot M}{8 \pi E^3} \,
   \frac{\zeta}{|\dot x_3 + \dot x_4|^2}
   \int_0^\infty d(\Delta t)
   \int_{\B^\xbx}
   \exp\bigl( - \tfrac12 |\dot M| \dot \Omega (B^\xbx)^2 \bigr)
\\ \times
   \frac{\B^\xbx}{(B^\xbx)^2}
   \cdot \grad_{\C_{12}^\xbx}
   \langle\C_{34}^\four,\C_{12}^\four,\Delta t|\C_{34}^\xbx,\C_{12}^\xbx,0\rangle
   \Bigr|_{\C_{12}^\xbx=0=\C_{34}^\four=\C_{12}^\four; ~ \C_{34}^\xbx=\B^\xbx}
  ,
\end {multline}
where
\begin {equation}
   \dot M = \dot x_3 \dot x_4 (\dot x_3{+}\dot x_4) E
\end {equation}
and
\begin {equation}
   \dot\Omega =
   \sqrt{
     -\frac{i\hat q_{\rm A}}{2E}
     \left(
       - \frac{1}{\dot x_3{+}\dot x_4}
       + \frac{1}{\dot x_4}
       + \frac{1}{\dot x_3}
     \right)
   } .
\end {equation}
\end {subequations}
(\ref{eq:yx4}) differs from the $4\bar y\bar x_2$ result (\ref{eq:4yx2})
only in (i) the change of $x_i$ to (\ref{eq:xmirror2b}),
(ii) the names used for some superscript labels, and
(iii) the transposition of the 4-particle propagator from
$\langle\C_{34},\C_{12},\Delta t|\C_{34}^\four,\C_{12}^\four,0\rangle$
to
$\langle\C_{34}^\four,\C_{12}^\four,\Delta t|\C_{34},\C_{12},0\rangle$.
The latter makes no difference to the form of the right-hand side
of eq.\ (\ref{eq:Cprop4}) for the propagator.%
\footnote{
  There are some other sign issues to worry about here, but they are
  resolved the same way as in appendix E.1 of
  ref.\ \cite{2brem}.
}
The only change that matters, then, is the change of $x_i$, as
asserted in the main text.


\section {Summary of Crossed and Sequential Formulas}
\label {app:crossseq}

For the sake of completeness, we thought it useful to include
a complete summary of all of the formulas necessary for a complete
evaluation of the total $\Delta\,d\Gamma/dx\,dy$ (\ref{eq:total}) in one
paper, especially since there have been corrections \cite{dimreg}
to the results of one of the earlier papers \cite{2brem}.  The
formulas for the contributions involving 4-gluon vertices have already
been given in the main text.  This appendix summarizes the
contributions from the crossed and sequential diagrams, as well as
giving some of the explicit lower-level formulas that were needed in
the main text.

It is possible to scale out the factors of $\hat q_{\rm A}$ and
$E$ from all of our numerical results by replacing
$\Delta t$ by the dimensionless variable
$\Delta\mathsf{t} \equiv (\hat q_{\rm A}/E)^{1/2} \Delta t$.
For numerics, it is convenient
to work in units where $\hat q_{\rm A}{=}1$ and $E{=}1$, which then gives
the result for
$\Delta\,d\Gamma/dx\,dy$ in units of $(\hat q_{\rm A}/E)^{1/2}$.


\subsection{Crossed Diagrams}

Here we collect the result for the crossed diagrams \cite{2brem}
as corrected by ref.\ \cite{dimreg}.  A brief summary of the
interpretation of each piece below can be found in section VIII of
ref.\ \cite{2brem}.

\begin {equation}
   \left[ \frac{d\Gamma}{dx\>dy} \right]_{\rm crossed}
   =
   A(x,y) + A(1{-}x{-}y,y) + A(x,1{-}x{-}y)
\label {eq:summary1}
\end {equation}
\begin {align}
   A(x,y) &=
   A^{\rm pole}(x,y)
   + \int_0^{+\infty} d(\Delta t) \> 
     2 \Re \bigl[ B(x,y,\Delta t) + B(y,x,\Delta t) \bigr]
\label {eq:summaryA}
\end {align}
\begin {align}
   B(x,y,\Delta t) &=
       C(\{\hat x_i\},\alpha,\beta,\gamma,\Delta t)
       + C(\{x'_i\},\beta,\alpha,\gamma,\Delta t)
       + C(\{\tilde x_i\},\gamma,\alpha,\beta,\Delta t)
\nonumber\\
   &=
       C({-}1,y,1{-}x{-}y,x,\alpha,\beta,\gamma,\Delta t)
       + C\bigl({-}(1{-}y),{-}y,1{-}x,x,\beta,\alpha,\gamma,\Delta t\bigr)
\nonumber\\ &\qquad\qquad
       + C\bigl({-}y,{-}(1{-}y),x,1{-}x,\gamma,\alpha,\beta,\Delta t\bigr)
\end {align}
\begin {equation}
   C = D - \lim_{\hat q\to 0} D
\label {eq:summaryC}
\end {equation}
\begin {align}
   D(x_1,&x_2,x_3,x_4,\alpha,\beta,\gamma,\Delta t) =
\nonumber\\ &
   \frac{\CA^2 \alphas^2 M_\ix M_\fx}{32\pi^4 E^2} \, 
   ({-}x_1 x_2 x_3 x_4)
   \Omega_+\Omega_- \csc(\Omega_+\Delta t) \csc(\Omega_-\Delta t)
\nonumber\\ &\times
   \Bigl\{
     (\beta Y_\bx Y_\Ax + \alpha \Ybar_{\bx\Ax} Y_{\bx\Ax}) I_0
     + (\alpha+\beta+2\gamma) Z_{\bx\Ax} I_1
\nonumber\\ &\quad
     + \bigl[
         (\alpha+\gamma) Y_\bx Y_\Ax
         + (\beta+\gamma) \Ybar_{\bx\Ax} Y_{\bx\Ax}
        \bigr] I_2
     - (\alpha+\beta+\gamma)
       (\Ybar_{\bx\Ax} Y_\Ax I_3 + Y_\bx Y_{\bx\Ax} I_4)
   \Bigl\}
\label {eq:summaryD}
\end {align}
\begin {align}
   A^{\rm pole}(x,y) &\equiv
   2\Re \Biggl[
    \frac{\CA^2 \alphas^2}{16\pi^2} \,
    x y (1{-}x)^2 (1{-}y)^2(1{-}x{-}y)^2
\nonumber\\ & \qquad~ \times
    \biggl\{
       -i
       [\Omega_{-1,1-x,x} + \Omega_{-(1-y),1-x-y,x}
        - \Omega_{-1,1-y,y}^* - \Omega_{-(1-x),1-x-y,y}^*]
\nonumber\\ & \qquad~ \qquad \times
       \biggl[
       \left(
         (\alpha + \beta)
         + \frac{(\alpha + \gamma) xy}{(1{-}x)(1{-}y)}
       \right) \ln \left( \frac{1{-}x{-}y}{(1{-}x)(1{-}y)} \right)
       + \frac{2(\alpha+\beta+\gamma) x y}{(1{-}x)(1{-}y)}
       \biggr]
\nonumber\\ &\qquad~ \quad
     - \pi
       [\Omega_{-1,1-x,x} + \Omega_{-(1-y),1-x-y,x}
        + \Omega_{-1,1-y,y}^* + \Omega_{-(1-x),1-x-y,y}^*]
\nonumber\\ & \qquad~ \qquad \times
       \left(
         (\alpha + \beta)
         + \frac{(\alpha + \gamma) xy}{(1{-}x)(1{-}y)}
       \right)
     \biggr\}
   \Biggr]
\end {align}
\begin {subequations}
\label {eq:I}
\begin {align}
   I_0 &=
   \frac{4\pi^2}{(X_\bx X_\Ax - X_{\bx\Ax}^2)}
\displaybreak[0]\\
   I_1 &=
   - \frac{2\pi^2}{X_{\bx\Ax}}
   \ln\left( 1 - \frac{X_{\bx\Ax}^2}{X_\bx X_\Ax} \right)
\displaybreak[0]\\
   I_2 &=
   \frac{2\pi^2}{X_{\bx\Ax}^2}
     \ln\left( 1 - \frac{X_{\bx\Ax}^2}{X_\bx X_\Ax} \right)
   + \frac{4\pi^2}{(X_\bx X_\Ax - X_{\bx\Ax}^2)}
\displaybreak[0]\\
   I_3 &=
   \frac{4\pi^2 X_{\bx\Ax}}{X_\Ax(X_\bx X_\Ax - X_{\bx\Ax}^2)}
\displaybreak[0]\\
   I_4 &=
   \frac{4\pi^2 X_{\bx\Ax}}{X_\bx(X_\bx X_\Ax - X_{\bx\Ax}^2)}
\end {align}
\end {subequations}
\begin {subequations}
\label {eq:XYZdef}
\begin {align}
   \begin{pmatrix} X_\bx & Y_\bx \\ Y_\bx & Z_\bx \end{pmatrix}
   &\equiv
   \begin{pmatrix} |M_\ix|\Omega_\ix & 0 \\ 0 & 0 \end{pmatrix}
     - i a_\bx^{-1\top} \uOmega \cot(\uOmega\,\Delta t)\, a_\bx^{-1}
\\
   \begin{pmatrix} X_\Ax & Y_\Ax \\ Y_\Ax & Z_\Ax \end{pmatrix}
   &\equiv
   \begin{pmatrix} |M_\fx|\Omega_\fx & 0 \\ 0 & 0 \end{pmatrix}
     - i a_\Ax^{-1\top} \uOmega \cot(\uOmega\,\Delta t)\, a_\Ax^{-1}
\\
   \begin{pmatrix} X_{\bx\Ax} & Y_{\bx\Ax} \\ \Ybar_{\bx\Ax} & Z_{\bx\Ax} \end{pmatrix}
   &\equiv
   - i a_\bx^{-1\top} \uOmega \csc(\uOmega\,\Delta t) \, a_\Ax^{-1}
\end {align}
\end {subequations}
\begin {equation}
   \uOmega \equiv \begin{pmatrix} \Omega_+ & \\ & \Omega_- \end{pmatrix}
\label {eq:uOmega}
\end {equation}
\begin {equation}
   M_\ix = x_1 x_4 (x_1{+}x_4) E ,
   \qquad
   M_\fx = x_3 x_4 (x_3{+}x_4) E
\end {equation}
\begin {equation}
   \Omega_\ix
   = \sqrt{ 
     -\frac{i \hat q_{\rm A}}{2E}
     \left( \frac{1}{x_1} 
            + \frac{1}{x_4} - \frac{1}{x_1{+}x_4} \right)
   } ,
   \qquad
   \Omega_\fx
   = \sqrt{ 
     -\frac{i \hat q_{\rm A}}{2E}
     \left( \frac{1}{x_3} + \frac{1}{x_4}
            - \frac{1}{x_3 + x_4}
     \right)
   }
\end {equation}
\begin {equation}
   a_\ybx =
   \begin{pmatrix} C^+_{34} & C^-_{34} \\ C^+_{12} & C^-_{12} \end{pmatrix}
\label {eq:af}
\end {equation}
\begin {equation}
   a_\yx =
   \frac{1}{(x_1+x_4)}
   \begin{pmatrix}
       -x_3 & -x_2 \\
       \phantom{-}x_4 &  \phantom{-}x_1
   \end {pmatrix}
   a_\ybx
\label {eq:ai}
\end {equation}
\begin {align}
   \begin{pmatrix} \alpha \\ \beta \\ \gamma \end{pmatrix}
   =
   \phantom{+}
   & \begin{pmatrix} - \\ + \\ + \end{pmatrix} \Biggl[
       \frac{x}{y^3(1{-}x)^3(1{-}y)^3(1{-}x{-}y)}
       + \frac{y}{x^3(1{-}x)^3(1{-}y)^3(1{-}x{-}y)}
\nonumber\displaybreak[3]\\ & \qquad\qquad
       + \frac{1{-}x}{x^3y^3(1{-}y)^3(1{-}x{-}y)}
       + \frac{1{-}y}{x^3y^3(1{-}x)^3(1{-}x{-}y)}
   \Biggr]
\nonumber\\
   + & \begin{pmatrix} + \\ - \\ + \end{pmatrix} \Biggl[
       \frac{x}{y^3(1{-}x)(1{-}y)(1{-}x{-}y)^3}
       + \frac{y}{x^3(1{-}x)(1{-}y)(1{-}x{-}y)^3}
\nonumber\displaybreak[3]\\ & \qquad\qquad
       + \frac{1{-}x{-}y}{x^3y^3(1{-}x)(1{-}y)}
       + \frac{1}{x^3y^3(1{-}x)(1{-}y)(1{-}x{-}y)^3}
   \Biggr]
\nonumber\\
   + & \begin{pmatrix} + \\ + \\ - \end{pmatrix} \Biggl[
        \frac{1{-}x}{xy(1{-}y)^3(1{-}x{-}y)^3}
        + \frac{1{-}y}{xy(1{-}x)^3(1{-}x{-}y)^3}
\nonumber\displaybreak[3]\\ & \qquad\qquad
        + \frac{1{-}x{-}y}{xy(1{-}x)^3(1{-}y)^3}
        + \frac{1}{xy(1{-}x)^3(1{-}y)^3(1{-}x{-}y)^3}
   \Biggr]
\label {eq:abc}
\end {align}
The $\hat q\to 0$ limit for the vacuum piece in (\ref{eq:summaryC})
corresponds to taking all $\Omega$'s to zero and so making the
replacements
\begin {equation}
   \Omega_\ix \to 0,
   \qquad
   \Omega_\fx \to 0,
   \qquad
   \uOmega \cot(\uOmega\,\Delta t) \to (\Delta t)^{-1} ,
   \qquad
   \uOmega \csc(\uOmega\,\Delta t) \to (\Delta t)^{-1} ,
\end {equation}
\begin {equation}
  \Omega_\pm \csc(\Omega_\pm \Delta t) \to (\Delta t)^{-1} .
\end {equation}


\subsection{4-particle frequencies and normal modes}
\label {app:modes}

Here we collect formulas for the large-$\Nc$ frequencies and normal modes
associated with 4-particle propagation (section V.B of ref.\ \cite{2brem}).

\begin {equation}
  \Omega_\pm =
  \left[ - \frac{i\hat q_{\rm A}}{4E} \left(
    \frac1{x_1} + \frac1{x_2} + \frac1{x_3} + \frac1{x_4} \pm \sqrt\Delta
  \right) \right]^{1/2}
\label {eq:Omegapm}
\end {equation}
\begin {equation}
  \Delta =
   \frac1{x_1^2} + \frac1{x_2^2} + \frac1{x_3^2} + \frac1{x_4^2}
   + \frac{(x_3{+}x_4)^2+(x_1{+}x_4)^2}{x_1 x_2 x_3 x_4}
\end {equation}
\begin {subequations}
\label {eq:Cmodes}
\begin {align}
   C^\pm_{34}
   &=
   \frac{x_2}{x_3+x_4} \sqrt{\frac{x_1 x_3}{2 N_\pm E}}
   \left[
      \frac1{x_3}-\frac1{x_1}+\frac1{x_4}+\frac{x_1}{x_3 x_2} \pm \sqrt\Delta
   \right]
\\
   C^\pm_{12}
   &=
   - \frac{x_4}{x_1+x_2} \sqrt{\frac{x_1 x_3}{2 N_\pm E}}
   \left[
      \frac1{x_1}-\frac1{x_3}+\frac1{x_2}+\frac{x_3}{x_1 x_4} \pm \sqrt\Delta
   \right]
\end {align}
\end {subequations}
\begin {equation}
   N_\pm \equiv
   - x_1 x_2 x_3 x_4 (x_1+x_3)\Delta
   \pm (x_1 x_4 + x_2 x_3)(x_1 x_2 + x_3 x_4) \sqrt\Delta
\label {eq:Npm}
\end {equation}


\subsection{Sequential Diagrams}

Here we collect the result for the sequential diagrams \cite{seq}.  A
brief summary of the interpretation of each piece below can be found
in section III of ref.\ \cite{seq}.  Symbols such as $\Omega_\pm$ or
$a_\yx$, which
are written in the exact same notation as symbols defined above,
are given by their definitions above.

\begin {align}
   \left[ \Delta \frac{d\Gamma}{dx\>dy} \right]_{\rm sequential}
   = \quad
   & {\cal A}_\seq(x,y) + {\cal A}_\seq(1{-}x{-}y,y) + {\cal A}_\seq(x,1{-}x{-}y)
\nonumber\\
   + ~ &
   {\cal A}_\seq(y,x) + {\cal A}_\seq(y,1{-}x{-}y) + {\cal A}_\seq(1{-}x{-}y,x)
\label {eq:dGammaseq}
\end {align}
\begin {equation}
   {\cal A}_\seq(x,y)
   =
   {\cal A}^{\rm pole}_\seq(x,y)
   + \int_0^{\infty} d(\Delta t) \>
     \Bigl[
        2 \Re \bigl( B_{\rm seq}(x,y,\Delta t) \bigr)
        + F_{\rm seq}(x,y,\Delta t)
     \bigr]
\label {eq:Aseq}
\end {equation}
\begin {align}
   B_\seq(x,y,\Delta t) &=
       C_\seq(\{\hat x_i\},\bar\alpha,\bar\beta,\bar\gamma,\Delta t)
\nonumber\\
   &=
       C_\seq({-}1,y,1{-}x{-}y,x,\bar\alpha,\bar\beta,\bar\gamma,\Delta t)
\end {align}
\begin {equation}
   C_\seq = D_\seq - \lim_{\hat q\to 0} D_\seq
\end {equation}
\begin {align}
   D_\seq(x_1,&x_2,x_3,x_4,\bar\alpha,\bar\beta,\bar\gamma,\Delta t) =
\nonumber\\ &
   \frac{\CA^2 \alphas^2 M_\ix M_\fx^\seq}{32\pi^4 E^2} \, 
   ({-}x_1 x_2 x_3 x_4)
   \Omega_+\Omega_- \csc(\Omega_+\Delta t) \csc(\Omega_-\Delta t)
\nonumber\\ &\times
   \Bigl\{
     (\bar\beta Y_\yx^\seq Y_\xbx^\seq
        + \bar\alpha \Ybar_{\yx\xbx}^{\,\seq} Y_{\yx\xbx}^\seq) I_0^\seq
     + (\bar\alpha+\bar\beta+2\bar\gamma) Z_{\yx\xbx}^\seq I_1^\seq
\nonumber\\ &\quad
     + \bigl[
         (\bar\alpha+\bar\gamma) Y_\yx^\seq Y_\xbx^\seq
         + (\bar\beta+\bar\gamma) \Ybar_{\yx\xbx}^{\,\seq} Y_{\yx\xbx}^\seq
        \bigr] I_2^\seq
\nonumber\\ &\quad
     - (\bar\alpha+\bar\beta+\bar\gamma)
       (\Ybar_{\yx\xbx}^{\,\seq} Y_\xbx^\seq I_3^\seq + Y_\yx^\seq Y_{\yx\xbx}^\seq I_4^\seq)
   \Bigl\}
\label {eq:Dseq}
\end {align}
\begin {align}
   F_\seq(x,y,\Delta t) =
   \frac{\alphas^2 P(x) P(\frac{y}{1-x})}{4\pi^2(1-x)}
   \Bigl[ &
      \Re(i\Omega_\ix) \,
      \Re\bigl( \Delta t \, (\Omega_\fx^\seq)^2
                \csc^2(\Omega_\fx^\seq \, \Delta t) \bigr)
\nonumber\\
      + &
      \Re(i\Omega_\fx^\seq) \,
      \Re\bigl( \Delta t \, \Omega_\ix^2 \csc^2(\Omega_\ix \, \Delta t) \bigr)
   \Bigr]
\label {eq:Fseq}
\end {align}
\begin {equation}
   {\cal A}_\seq^{\rm pole}(x,y)
   = \frac{\alphas^2 \, P(x) \, P(\yfrak)}{2\pi^2(1-x)}
   \biggl(
      - \tfrac12
           \Re(i \Omega_{E,x} + i \Omega_{(1-x)E,\yfrak})
      + \tfrac{\pi}{4}
           \Re(\Omega_{E,x} + \Omega_{(1-x)E,\yfrak})
   \biggr)
\end {equation}
\begin {subequations}
\label {eq:Iseq}
\begin {align}
   I_0^\seq &=
   \frac{4\pi^2}{[X_\yx^\seq X_\xbx^\seq - (X_{\yx\xbx}^\seq)^2]}
\displaybreak[0]\\
   I_1^\seq &=
   - \frac{2\pi^2}{X_{\yx\xbx}^\seq}
   \ln\left( 1 - \frac{(X_{\yx\xbx}^\seq)^2}{X_\yx^\seq X_\xbx^\seq} \right)
\displaybreak[0]\\
   I_2^\seq &=
   \frac{2\pi^2}{(X_{\yx\xbx}^\seq)^2}
     \ln\left( 1 - \frac{(X_{\yx\xbx}^\seq)^2}{X_\yx^\seq X_\xbx^\seq} \right)
   + \frac{4\pi^2}{[X_\yx^\seq X_\xbx^\seq - (X_{\yx\xbx}^\seq)^2]}
\displaybreak[0]\\
   I_3^\seq &=
   \frac{4\pi^2 X_{\yx\xbx}^\seq}
        {X_\xbx^\seq[X_\yx^\seq X_\xbx^\seq - (X_{\yx\xbx}^\seq)^2]}
\displaybreak[0]\\
   I_4^\seq &=
   \frac{4\pi^2 X_{\yx\xbx}^\seq}
        {X_\yx^\seq[X_\yx^\seq X_\xbx^\seq - (X_{\yx\xbx}^\seq)^2]}
\end {align}
\end {subequations}
\begin {subequations}
\label {eq:XYZseqdef}
\begin {align}
   \begin{pmatrix} X_\yx^\seq & Y_\yx^\seq \\ Y_\yx^\seq & Z_\yx^\seq \end{pmatrix}
   &\equiv
   \begin{pmatrix} |M_\ix|\Omega_\ix & 0 \\ 0 & 0 \end{pmatrix}
     - i a_\yx^{-1\top} \uOmega \cot(\uOmega\,\Delta t)\, a_\yx^{-1}
   =
   \begin{pmatrix} X_\yx & Y_\yx \\ Y_\yx & Z_\yx \end{pmatrix} ,
\\
   \begin{pmatrix} X_\xbx^\seq & Y_\xbx^\seq \\ Y_\xbx^\seq & Z_\xbx^\seq \end{pmatrix}
   &\equiv
   \begin{pmatrix} |M_\fx^\seq|\Omega_\fx^\seq & 0 \\ 0 & 0 \end{pmatrix}
     - i (a_\xbx^\seq)^{-1\top} \uOmega \cot(\uOmega\,\Delta t)\, (a_\xbx^\seq)^{-1} ,
\\
   \begin{pmatrix} X_{\yx\xbx}^\seq & Y_{\yx\xx}^\seq \\
                   \Ybar_{\yx\xbx}^\seq & Z_{\yx\xbx}^\seq \end{pmatrix}
   &\equiv
   - i a_\yx^{-1\top} \uOmega \csc(\uOmega\,\Delta t) \, (a_\xbx^\seq)^{-1}
\end {align}
\end {subequations}
\begin {equation}
   M_\fx^\seq = x_2 x_3 (x_2{+}x_3) E
\label {eq:Mfseq}
\end {equation}
\begin {equation}
   \Omega_\fx^\seq
   = \sqrt{ 
     -\frac{i \hat q_{\rm A}}{2E}
     \left( \frac{1}{x_2} + \frac{1}{x_3}
            - \frac{1}{x_2 + x_3}
     \right)
   }
\end {equation}
\begin {equation}
   a_\xbx^\seq \equiv 
   \begin{pmatrix} 0 & 1 \\ 1 & 0 \end{pmatrix} a_\yx
\end {equation}
\begin {equation}
   P(x) = P_{g\to gg}(x) =
   \CA \, \frac{[1 + x^4 + (1{-}x)^4]}{x (1{-}x)}
\label {eq:Pggg}
\end {equation}
\begin {align}
   \begin{pmatrix}
      \bar\alpha \\ \bar\beta \\ \bar\gamma
   \end{pmatrix}_{x\bar y\bar x y}
   =
   \phantom{+}
   & \begin{pmatrix} - \\ + \\ + \end{pmatrix}
       \frac{4}{xy(1{-}x)^6(1{-}x{-}y)}
\nonumber\\
   + & \begin{pmatrix} + \\ - \\ + \end{pmatrix} \Biggl[
       \frac{1}{x^3y^3(1{-}x)^2(1{-}x{-}y)^3}
       + \frac{1{-}x{-}y}{x^3y^3(1{-}x)^2}
\nonumber\\ & \qquad\qquad
       + \frac{x}{y^3(1{-}x)^2(1{-}x{-}y)^3}
       + \frac{y}{x^3(1{-}x)^2(1{-}x{-}y)^3}
   \Biggr]
\nonumber\\
   + & \begin{pmatrix} + \\ + \\ - \end{pmatrix} \Biggl[
       \frac{(1{-}x)^2}{x^3y^3(1{-}x{-}y)^3}
       + \frac{(1{-}x{-}y)}{x^3y^3(1{-}x)^6}
       + \frac{x(1{-}x{-}y)}{y^3(1{-}x)^6}
\nonumber\\ & \qquad\qquad
       + \frac{y}{x^3(1{-}x)^6(1{-}x{-}y)^3}
       + \frac{xy}{(1{-}x)^6(1{-}x{-}y)^3}
   \Biggr]
\label {eq:abcbar}
\end {align}


\section {Approximate analytic formula fitted to result}
\label {app:approx}

Similar to what was done in Appendix A of ref.\ \cite{seq},
the following approximation reproduces the results of fig.\
\ref{fig:4pointResult} with a maximum absolute error%
\footnote{
  We quote absolute error rather than relative error because the result
  is zero along the red curve in fig.\ \ref{fig:4pointResult}.
  Any numerical approximation will have infinite relative error exactly
  on this curve, which is irrelevant to the question of how useful the
  approximation is.
}
of 0.017 for
all $y > 10^{-4}$ (assuming one permutes the final state gluons to choose
$y < x < z$, just as in fig.\ \ref{fig:4pointResult}):

\begin{equation}
  \pi^2 \,x \,y^{\frac{3}{2}} \,\Delta\,\frac{d \Gamma}{dx dy} =
  \sum^3_{m=0}\sum^4_{n=0}
  \left(a_{m n} + b_{m n} \Big(\frac{y}{x}\Big)^{\frac{1}{3}} \right) s^m t^n ,
\label {eq:approx}
\end {equation}
where the parameters
\begin {equation}
  s \equiv \frac{2 (x - y)}{t},
\qquad
  t \equiv 2 x + y
\end{equation}
each vary independently from 0 to 1.
The numerical coefficients $a_{mn}$ and $b_{mn}$ are given
in tables \ref{tab:amn} and \ref{tab:bmn}.  We have made no
effort to make the approximation work well for $y < 10^{-4}$.

\begin{table}[t]
\begin{tabular}{|c||*{5}{c|}}\hline
\backslashbox{m}{n}
&\makebox[2em]{0}&\makebox[2em]{1}&\makebox[2em]{2}
&\makebox[2em]{3}&\makebox[2em]{4}\\\hline\hline
0& -5.00370& 41.0019& -200.721& 355.883& -204.864\\ \hline
 1& 6.37665& -82.3722& 414.714& -739.307& 424.729\\ \hline
 2& -2.34616& 49.6745& -253.978& 453.977& -260.422\\ \hline
 3& 0.0251252& -7.35668& 38.8566& -69.7090& 40.0310\\ \hline
\end{tabular}
\caption{
  \label{tab:amn}
  The coefficients $a_{mn}$ in eq.\ (\ref{eq:approx}).
}
\end{table}

\begin{table}[t]
\begin{tabular}{|c||*{5}{c|}}\hline
\backslashbox{m}{n}
&\makebox[2em]{0}&\makebox[2em]{1}&\makebox[2em]{2}
&\makebox[2em]{3}&\makebox[2em]{4}\\\hline\hline
0& 5.48414& -41.2208& 201.848& -357.473& 206.179\\ \hline
 1& -3.83142& 62.2511& -316.542& 565.450& -325.181\\ \hline
 2& 0.238156& -19.3169& 101.583& -182.650& 105.175\\ \hline
 3& 0.401059& -3.48365& 16.8782& -29.6769& 16.7608\\ \hline
\end{tabular}
\caption{
  \label{tab:bmn}
  The coefficients $b_{mn}$ in eq.\ (\ref{eq:approx}).
}
\end{table}


\end {document}